\documentclass[twoside]{article}

\usepackage[accepted]{aistats2026}
\usepackage[round]{natbib}

\usepackage{xcolor}
\definecolor{red}{rgb}{0.0,0.0,0.0}
\usepackage[hidelinks]{hyperref}
\hypersetup{
    colorlinks,
    linkcolor={blue!50!black},
    citecolor={blue!50!black},
    urlcolor={blue!50!black}
}
\usepackage{adjustbox}
\usepackage{url}
\usepackage{subcaption}
\usepackage{booktabs}
\usepackage[colorinlistoftodos,prependcaption,textsize=tiny]{todonotes}

\usepackage{amsmath, amssymb, amsthm, bbm, mathtools}
\usepackage{algorithm, algpseudocode}
\def\h{\mathbf{h}}
\def\v{\mathbf{v}}
\def\x{\mathbf{x}}
\def\y{\mathbf{y}}
\def\P{{\mathbb{P} }}
\def\E{{\mathbb{E} }}

\numberwithin{equation}{section}
\theoremstyle{plain}
\newtheorem{theorem}{Theorem}

\newtheorem{corollary}{Corollary}

\newtheorem{proposition}{Proposition}
\newcommand{\norm}[1]{\left\lVert#1\right\rVert}

\begin{document}

%
\runningtitle{MCMC for Intractable Probabilistic Graphical Models}

%

\twocolumn[

\aistatstitle{Exact and Approximate MCMC for Doubly-intractable Probabilistic Graphical Models Leveraging the Underlying Independence Model}

\aistatsauthor{ Yujie Chen \And Antik Chakraborty \And  Anindya Bhadra }

\aistatsaddress{Department of Statistics \\Purdue University\\ chen1866@purdue.edu \And Department of Statistics \\Purdue University\\ antik015@purdue.edu \And Department of Statistics \\Purdue University\\ bhadra@purdue.edu} ]

\begin{abstract}
 Bayesian inference for doubly-intractable {\color{red} pairwise exponential} graphical models typically involves variations of the exchange algorithm or approximate Markov chain Monte Carlo (MCMC) samplers. However, existing methods for both classes of algorithms require either perfect samplers or sequential samplers for complex models, which are often either not available, or suffer from poor mixing, especially in high dimensions. We develop a method that does not require perfect or sequential sampling, and can be applied to both classes of methods: exact and approximate MCMC. The key to our approach is to utilize the \emph{tractable independence model} underlying the \emph{intractable probabilistic graphical model} for the purpose of constructing a finite sample unbiased Monte Carlo (and \emph{not} MCMC) estimate of the Metropolis--Hastings ratio. This innovation turns out to be crucial for scalability in high dimensions. The method is demonstrated on the Ising model. Gradient-based alternatives to construct a proposal, such as Langevin and Hamiltonian Monte Carlo approaches, also arise as a natural corollary to our general procedure, and are demonstrated as well. 
\end{abstract}

\section{\uppercase{Introduction}}
{\color{red} Undirected graphical models \citep{koller2009probabilistic}, e.g. Markov Random Fields,} are a widely popular tool to describe joint distributions of a set of random variables through conditional dependencies. Given a set of random variables $\{X_1, \ldots, X_p\}$, these models have a joint distribution of the following form:
\begin{equation}
    p(\x; \theta) = \dfrac{f(\x;\theta)}{z(\theta)} ,\; \x = (x_1, \ldots, x_p), \label{eq:basic}
\end{equation}
where the parameter $\theta \in \mathbb{R}^{p \times p} $ encodes the strength of conditional dependencies between the variables, and $z(\theta)$ is a normalizing constant. {\color{red} A powerful subclass of these models is the so-called pairwise exponential family graphical models (PEGMs), where $p(\x;\theta)$ defines an exponential family model involving linear and pairwise interaction terms in $\x$ parametrized by a symmetric matrix $\theta \in \mathbb{R}^{p \times p}$. Examples include the multivariate Gaussian or the Ising model, with a wide range of applications.}
{\color{red} Within this class, only the multivariate Gaussian admits a tractable normalizing constant $z(\theta)$. For other models, $z(\theta)$ is intractable. }To see why this is the case, one may take a concrete example of the Ising model \citep{ising1924beitrag}, for which \eqref{eq:basic} reads:
\begin{equation}
    p(\x; \theta) = \dfrac{\exp\left\lbrace\sum_j x_j \theta_{jj} + \\ \sum_{jk}x_jx_k \theta_{jk}\right\rbrace}{z(\theta)} , \label{eq:basic2}
\end{equation}
where $x_j\in\{0,1\}$ and $\theta_{jk}\in \mathbb{R}$. Clearly, in order to compute $z(\theta)$, one must sum the numerator over $2^p$ possible configurations of $\x$, which leads to an exponential complexity combinatorial problem. Other models where the same issue arises include the Potts model \citep{potts1952some}, the Poisson graphical model \citep{besag1974spatial} and many others. The intractability of $z(\theta)$ poses a critical challenge in conducting standard statistical inference, including Bayesian inference, which is the focus of this work. Bayesian inference proceeds by eliciting a prior $\pi(\theta)$ on $\theta$. By Bayes theorem, one then obtains the posterior $\pi(\theta \mid \x)$.
However, standard MCMC methods cannot be applied for posterior sampling in this case. For example, the most general MCMC procedure, the Metropolis-Hastings (M--H) algorithm, requires a new proposed state  $\theta' \sim q(\cdot \mid \theta)$ conditional on the current state $\theta$ of the Markov chain. This new state is accepted with probability $\alpha_{MH}(\theta, \theta')=\min{\{R_{MH}(\theta, \theta'), 1\}}$ where: 
\begin{align}\label{eq:alpha_MH}
R_{MH}(\theta, \theta') = \dfrac{f(\x; \theta') \pi(\theta') q(\theta \mid \theta') z(\theta) }{f(\x;\theta) \pi(\theta) q(\theta' \mid \theta)z(\theta')}.
\end{align}
Hence, for models with intractable $z(\theta)$, computing the above acceptance probability is not possible analytically. This also applies to partially observed models. But nevertheless, valid posterior sampling can still be executed following one of the two broadly general strategies, which are discussed below.

\subsection{Related Works in Intractable Models}

{\bf Exact MCMC methods:} Canonical exact MCMC approaches consist of the auxiliary variable method of \cite{moller2006efficient}, and its generalization, the exchange algorithm \citep{murray2012mcmc}. These methods consider sampling from an augmented posterior $\pi(\y, \theta \mid \x)$ with the desired posterior $\pi(\theta \mid \x)$ as its marginal, where the state space of $\y$ is the same as $\x$. An M--H proposal $q(\y', \theta' \mid \y, \theta)$ is considered to move the chain to a new state. This proposal is constructed carefully to bypass the evaluation of $z(\theta)$. The second strategy involves the pseudo-marginal MCMC approach \citep{andrieu2009pseudo}, wherein an unbiased estimator of the likelihood is constructed at every step of the chain. 

Auxiliary variable methods, while appealing, lack flexibility, in that their validity relies heavily on the ability to perform exact sampling from $p(\y \mid \theta)$, which is often not feasible in practice. On the other hand, implementing an exact/approximate pseudo-marginal approach also requires an unbiased estimate of the inverse of the normalizing constant. This is typically done by using the sum-estimator \citep{lyne2015russian}. An implicit assumption here is that an unbiased estimator of the normalizing constant is readily available. For example, \cite{lyne2015russian} use a sequential Monte Carlo sampler to construct an unbiased estimate of the normalizing constant, which when implemented inside an MCMC chain, could become prohibitive.

{\bf Approximate MCMC methods:} Parallel to the exact MCMC methods, there exists a strand of works that can be broadly classified as approximate MCMC. These methods are not pseudo-marginal approaches in the strictest sense, i.e., they do not target an augmented posterior, but rather, try to approximate the M--H acceptance ratio in some sense. Common approaches include
the approximate algorithm by \citet{atchade2013bayesian}, noisy MCMC \citep{alquier2016noisy}, double MH \citep{liang2010double}, and noisy Hamiltonian MCMC \citep{stoehr2019noisy}. A common framework of theoretical justification for these methods can be found in \citet{alquier2016noisy}. 

\subsection{Key Intuition Behind the Current Work}
While the model of \eqref{eq:basic2} is indeed intractable, there is one specific configuration for which the model is, in fact, tractable. Take each $\theta_{jk}=0,\, j\ne k$, and denote this parameter by $\phi=\mathrm{diag}(\theta)$, i.e., $\phi$ merely strips out the diagonal elements of $\theta$ and zeros out the off-diagonals. Then, \eqref{eq:basic2} reads:
\begin{equation}
    p(\x; \phi) = \dfrac{\exp\left\lbrace\sum_j x_j \theta_{jj}\right\rbrace}{z(\phi)} =\dfrac{\prod_j\exp\left\lbrace x_j \theta_{jj}\right\rbrace}{z(\phi)}. \label{eq:basic3}
\end{equation}
This is the \emph{independence model} underlying the general model, as can be seen from the product factorization of \eqref{eq:basic3}, and $z(\phi)$ can now be obtained by $p$ univariate marginalizations, which requires considering $2p$ configurations, and not $2^p$. The other relevant feature is that it is trivial to sample from $p(\x;\phi)$; one only needs to draw $p$ independent Bernoulli variables in batch. To handle the general case, we show in the rest of the paper how this important special case can be leveraged via importance sampling.

\subsection{Summary of Our Contributions}
\begin{enumerate}
    \item We provide an exact pseudo-marginal MCMC approach for intractable PEGMs that leaves the target posterior invariant.
    \item Unlike existing pseudo-marginal approaches, or double MH approaches, our method does not require exact sampling from {\color{red}$p(\cdot;\theta),$} which is computationally prohibitive and impractical in high dimensions. This is done exploiting the \emph{independence model} underlying an \emph{intractable model}.
    \item We also develop an approximate MCMC method and study its properties.
    \item Numerical demonstrations show the pseudo-marginal approach has better mixing properties compared to the exchange algorithm, and especially in high dimensions, the approximate sampler is as good as the exchange algorithm.
\end{enumerate}

\section{\uppercase{Background}}

\subsection{Pairwise Exponential Family Graphical Models}
The models considered here are parameterized by a graph $G = (V, E)$, where $V = \{X_1, \ldots, X_p\}$ is the set of vertices/random variables and $E$ is the set of edges between the vertices. We shall focus on undirected graphical models, i.e. if $ (j, k)\in E$ then $(k,j) \in E$. Among these models, the pairwise exponential family graphical models (PEGM) is particularly well-studied as it has simple exponential family conditional distributions for each variable in $V$. For this subclass of models, a parameter $\theta \in \mathbb{R}^{p \times p}$ encodes the graph, noting that $\theta_{jk} \neq 0 $ iff $(j, k) \in E$. The joint distribution of a PEGM has the form of~\eqref{eq:basic} where $f(\x;\theta) = \exp\left\lbrace\sum_j T(x_j) \theta_{jj} + \\ \sum_{jk}T(x_j, x_k) \theta_{jk}\right\rbrace$. Here, $T(x_j)$ and $T(x_j, x_k)$ are the sufficient statistics of the model and $z(\theta) = \int_\x f(\x; \theta)d\x$, where the integral is taken with respect to an appropriate dominating measure. The parameter space $\Theta$ is such that $\int f(\x;\theta) d\x < \infty$. It is known that $\Theta$ is convex \citep{wainwright2008graphical}.
A standard (tractable) example is the Gaussian graphical model, where $\theta$ is the inverse covariance matrix, $T(x_j) = x_j^2$, $T(x_j, x_k) = x_j x_k$, and $z(\theta)  = |\theta|^{-1/2}$. Here, $\theta \in \Theta$, with $\Theta$ being the space of $p \times p$ positive definite matrices. Moreover, any variable conditional on the rest, i.e. $X_j \mid X_{-j}$, is a univariate Gaussian. However, in general, $z(\theta)$ is intractable.

For PEGMs, the distribution $X_j \mid X_{-j}$ equivalently determines the joint distribution of the variables via Brook's lemma \citep[see, e.g.,][]{brook1964distinction,besag1974spatial}. Indeed, when {\color{red} $X_j \mid X_{-j} \sim \text{Bernoulli}(\mathrm{expit}(\theta_{jj} + 2 \sum_{k\ne j} \theta_{jk} x_k))$} for all $j = 1, \ldots, p$, then the joint model is the familiar Ising model. For the Ising model, $\Theta$ is the set of all $p\times p$ matrices. Other examples include the Poisson graphical model \citep{besag1974spatial, yang2013poisson} and the Potts model \citep{potts1952some}.  

It is also possible to consider \emph{partially observed} PEGMs. Suppose $\x = (\v, \h)$ and the joint distribution of \emph{visible} ($\v$) and \emph{hidden} ($\h$) variables is Ising with parameter $\theta \in \mathbb{R}^{p\times p}$. Consider a special case where $p_\theta(\v \mid \h) = \prod_{j=1}^{m_1} p_\theta(v_j \mid \h)$, and $p_\theta(\h \mid \v) = \prod_{k=1}^{m_2} p_\theta(h_k \mid \v)$. In other words, the visible variables are conditionally independent given the hidden variables and vice versa. The resulting distribution of the visible variables from this joint model is known as the Restricted Boltzmann machine or RBM \citep{salakhutdinov2007restricted}. The restriction refers to the conditional independence structure of the model. When no such independence is allowed, the distribution of the visible variables is known as a Boltzmann machine or BM \citep{hinton2007boltzmann}. Since exponential family is closed under conditioning, but not necessarily closed under marginalization \citep{barndorff2014information}, these models allow $p_\theta(\v ) = \int p_\theta(\v, \h) d\h$ to depart from exponential family to capture more complex dependence, while still allowing for methods such as contrastive divergence \citep{hinton2002training} to be used for training.

\subsection{Pseudo-marginal MCMC}
Consider sampling from $\pi(\theta \mid \x) \propto [f(\x;\theta)/z(\theta)]\pi(\theta)$. Due to the intractability of $z(\theta)$, the M--H acceptance probability cannot be computed. However, let $\hat{p}(\x;\theta \mid u)$ be an unbiased Monte Carlo estimator of $p(\x;\theta)$ where $u \sim p(u)$, i.e. $\int \hat{p}(\x;\theta\mid u) p(u) du = p(\x; \theta)$ for every $\x$ and $\theta$. {\color{red} The corresponding estimate of the posterior of $\theta$ is $\hat{\pi}(\theta\mid \x, u) = \hat{p}(\x;\theta \mid u) \pi(\theta)/p(\x),$ where $p(\x)$ is the marginal distribution of the data, i.e., $p(\x) = \int p(\x;\theta)\pi(\theta) d\theta$. Set $\hat{\pi}(\theta, u \mid \x) = \hat{\pi}(\theta \mid \x, u) p (u)$.} By construction, this joint distribution over $(\theta, u)$ has $\pi(\theta \mid \x)$ as marginal over $\theta$. Now consider an M--H sampler for $\hat{\pi}(\theta, u\mid \x)$ with proposal distribution $q(\theta'\mid \theta) p(u')$. Then the resulting acceptance ratio is:
\begin{align*}
 &\alpha_{PM}(\theta, \theta')=\min \left\lbrace \dfrac{\hat{\pi}(\theta' \mid \x, u') p(u') q(\theta \mid \theta') p(u)}{\hat{\pi}(\theta \mid \x, u) p(u) q(\theta' \mid \theta) p(u')},\; 1 \right\rbrace.
\end{align*}
Importantly, all terms in $\alpha_{PM}(\theta, \theta')$ are computable. Moreover, the chain has  $\pi(\theta \mid \x)$ as the marginal over $\theta$ at stationarity. This procedure is known as the pseudo-marginal MCMC \citep{andrieu2009pseudo} (PM-MCMC). For a successful implementation in the present context, one needs an unbiased estimator of $1/z(\theta)$ which is positive.
When $n$ independent copies of $X$ are observed, we need an unbiased estimator of $[z(\theta)]^{-n}$. Note that if $T$ is unbiased for $z(\theta)$, i.e., $\mathbb{E}(T)=z(\theta)$, then, in general, $\mathbb{E}(T^{-1})\ne [z(\theta)]^{-1}$.

\subsection{The Exchange Algorithm}
{\color{red} A valid Markov chain targeting $\pi(\theta\mid \x)$ can also be developed by constructing an unbiased estimator of the M--H ratio. }
Recall from \eqref{eq:alpha_MH} that the M--H ratio involves $z(\theta)/z(\theta')$. The exchange algorithm  \citep{murray2012mcmc} is an auxiliary variable method where $z(\theta)/z(\theta')$ is unbiasedly estimated by $f(W;\theta)/f(W;\theta')$ with $W \sim p(\cdot;\theta')$. It is easy to see that $\E_W[f(W;\theta)/f(W;\theta')] = z(\theta)/z(\theta')$. With $n$ independent realizations of $W$, the exchange algorithm sets {\color{red} $\alpha_{EX}(\theta,\theta')=\min\{{R}_{EX}(\theta,\theta'),1\},$} with,
{\color{red} 
\small
\begin{equation}
    {R}_{EX}(\theta,\theta')=\dfrac{\prod_{l=1}^n f(\x_l;\theta') \pi(\theta') q(\theta\mid \theta')}{\prod_{l=1}^n f(\x_l;\theta) \pi(\theta) q(\theta'\mid \theta)} \dfrac{\prod_{l=1}^n f(\mathbf{w}_l;\theta)}{\prod_{l=1}^n f(\mathbf{w}_l;\theta')},
\end{equation}
\normalsize
}
where $w_l \overset{iid}{\sim} p(\cdot;\theta'), \, l = 1, \ldots, n$. Clearly, $\E_W[R_{EX}(\theta, \theta')] = R_{MH}(\theta, \theta')$. However, this remarkably simple workaround to \emph{cancel out} the intractable $[z(\theta)/z(\theta')]^{-n}$ disguises some key underlying assumptions that can be inherently limiting, outlined below.
\begin{enumerate}
    \item  It is assumed that sampling $w_l \sim p(\cdot;\theta')$ is possible, and the number of auxiliary variables drawn is equal to $n$, the number of observed samples. Although perfect samplers \citep{propp1996exact} exist to address the first concern, implementing them in high dimensions is computationally prohibitive, especially if $n$ is large. In practice, one often resorts to a Gibbs sampler to simulate the auxiliary data, as in double MH \citep{liang2010double}, which destroys the theoretical validity of the exchange algorithm.
    
    \item More crucially, the fact that the number of auxiliary samples $N$ has to be exactly equal to the number of observed samples $n$ imposes some artificial bottleneck on controlling the variance of the estimates.  It is of interest to decouple $N$ and $n$.
    \end{enumerate}

\section{\uppercase{Exact mcmc using an unbiased estimate of the likelihood}}
In this section, we develop an unbiased estimator of the likelihood function {\color{red}akin to \cite{lyne2015russian, chopin2024towards}}, which can be used to conduct MCMC. Suppose $n$ i.i.d. copies of $X$ are available, i.e. $\x_l \overset{iid}{\sim} p(\cdot;\theta)$, $l = 1, \ldots, n$, and $\theta \sim \pi(\theta)$ is some prior density over $\theta$. We assume that $\pi(\theta)$ can be evaluated analytically for every $\theta \in \Theta$. Set $\mathcal{D} = \{\x_1, \ldots, \x_n\}$. The posterior density of $\theta$ is:
\begin{equation}
    \pi(\theta \mid \mathcal{D}) \propto \left[\prod_{l=1}^n f(\x_l;\theta)\right] [z(\theta)]^{-n} \pi(\theta).
\end{equation}
To construct a valid pseudo-marginal algorithm, we then need an unbiased estimate of $[z(\theta)]^{-n}$. Suppose $\mu = z(\theta)/z(\phi)$, where $\phi = \text{diag}(\theta)$. Then for a suitably chosen $\nu$,
\begin{align*}
    [z(\theta)]^{-n} = \left[\frac{\nu}{z(\phi)}\right]^{n} &\{1 - (1 - \nu \mu)\}^{-n}\\
     = \left[\frac{\nu}{z(\phi)}\right]^{n} \sum_{k=0}^\infty &\gamma_k (1 - \nu\mu)^{k}\\
     = \left[\frac{\nu}{z(\phi)}\right]^{n}  g_\nu(\mu)&, 
\end{align*}
for $g_\nu(\mu)=\sum_{k=0}^\infty \gamma_k (1 - \nu\mu)^{k}$ {\color{red} and 
$\,\gamma_k = \binom{n+k-1}{k}.$} This {\color{red} Taylor expansion of} $g_\nu(\mu)$ is convergent if and only if  $|1 - \nu \mu| < 1$. We shall treat $\nu = \nu(\theta)$ as a tuning parameter, and discuss how we choose $\nu$ later. Crucially, in the above formulation, $z(\phi)$ is explicitly known as it corresponds to the normalizing constant of an independent PEGM. 

We can now attempt to estimate $g_\nu(\mu)$. One possibility is that we draw a random non-negative integer from some distribution and truncate the sum to our sampled value. Let this random variable be $R$. Define:
{\color{red} 
\begin{align*}
    T^\star = \sum_{k=0}^R \dfrac{\gamma_k}{\P(R\geq k)} \left( 1 - \nu \mu\right)^k.
\end{align*}
}
Then,
{\color{red} 
\begin{align*}
    \E(T^\star) &= \sum_{r=0}^\infty \left[ \sum_{k=0}^r \dfrac{\gamma_k}{\P(R\geq k)} \left( 1- \nu \mu \right)^k\right] \P(R = r) \\
    & = \sum_{k=0}^\infty \dfrac{\gamma_k}{\P(R\geq k)} \left( 1 - \nu\mu\right)^k \sum_{r \geq k} \P(R = r) \\
    & = g_\nu(\mu).
\end{align*}
}
The interchange of sums in the previous display is feasible due to Fubini's theorem and the fact that $|1 - \nu \mu| < 1$. {\color{red} We note here that this estimator only takes care of the infinite sum in $g_{\nu}(\mu)$ since it involves the unknown quantity $\mu$. }To complete the specification of the unbiased estimator, we need an unbiased estimate of $(1 - \nu \mu)^k$ for $k = 0, 1, \ldots$, or more specifically, $\mu$. Set 
\begin{equation}
    \widetilde{T} = \widetilde{T}(\theta) = \frac{1}{N} \sum_{i=1}^N \frac{f(\y_i;\theta)}{f(\y_i;\phi)}, \quad \y_i \overset{iid}{\sim }p(\cdot;\phi).
\end{equation}
Clearly, $\widetilde{T}$ is an unbiased estimator of $\mu = z(\theta)/z(\phi)$. Indeed,
$$\E_{Y \sim p(\cdot; \phi)}\left[ \dfrac{f(Y;\theta)}{f(Y; \phi)}\right] = \int \left[ \dfrac{f(y;\theta)}{f(y; \phi)}\right] p(y;\phi) dy= \mu.$$
Under very mild conditions, this estimator has finite variance \citep[][Proposition 3.2]{chen2024likelihood}. Moreover, sampling $\y \sim p(\cdot;\phi)$ can be done in batches since $\phi$ represents the independence model. With independent copies of $\widetilde{T}$, define for $r = 0, 1, \ldots,$
$${\color{red} U_{r,k} = \prod_{j=1}^k (1 - \nu \widetilde{T}_j), \quad 0< k \leq r.}$$
Next, we can define the estimator:
\begin{align*}
    T = \sum_{k=0}^R \dfrac{\gamma_k}{\P(R\geq k)} U_{R,k}.
\end{align*}
Suppose $U_{R,k}$ is independent of $R$. By definition, $\E(U_{R,k}) = (1 - \nu \mu)^k$. Thus,
\begin{align*}
    \E(T) &= \E_{(R,U)} \left[ \sum_{k=0}^R \dfrac{\gamma_k}{\P(R\geq k)} U_{R,k} \right]
     = g_\nu(\mu).
\end{align*}
This expectation is well-defined if $\E(|T|)$ exists. Two conditions ensure this. First, $a_k = \sup_{r \geq k} \E [|U_{r,k}|] < \infty$, and second, $\sum_{k=0}^\infty |\gamma_k| a_k < \infty$. We next show these conditions are true under mild assumptions. 
\begin{proposition}\label{prop:finite_absolute_expectation}
   Suppose $\nu$ is such that $\E|1 - \nu \widetilde{T}|< 1$. Then $\E(|T|)$ is finite.
\end{proposition}
All technical proofs can be found in Supplementary Section~\ref{sec:supp_proofs}. {\color{red} As mentioned at the beginning of the section,} the development {\color{red}until} this point is similar to other sum-based estimators of smooth functions such as \citet{lyne2015russian} and \citet{chopin2024towards}. For these estimators, a point of expansion of the infinite series is required, which is a tuning parameter for the method. The key difference between the proposed method and those previous approaches is that we expand $(1 - x)^{-n}$ around 0 where $x = (1 - \nu \mu)$. The parameter $\nu$ plays the same role in our case. {\color{red} Moreover,} these methods typically assume an unbiased estimator of $\mu$ is readily available, and often use expensive sequential Monte Carlo techniques to construct such estimators. Here, we explicitly provide a finite-variance estimator which can be constructed avoiding sequential samplers altogether. Additionally, \citet[][Proposition 3.4]{chen2024likelihood} show that to obtain reliable estimates of $\mu$, the number of importance samples $N$ for sparse high-dimensional PEGMs needs to scale as: {\color{red} $N = O(p)$}, reflecting a modest computational demand for our approach.

\subsection{Variance of $T$}
The choice of the distribution of the random truncation variable $R$ plays a significant role in establishing properties of the variance.
Due to the law of total variance, we have the decomposition: $\text{\emph{var}}(T) = \E[\text{\emph{var}}(T\mid R)] + \text{\emph{var}}[\E(T \mid R)].$ This decomposition is instructive, as the first term captures variation due to the unbiased estimates of $\mu$, whereas the second term captures the variation due to the random truncation. In Theorem~\ref{thm:variance}, we bound these two terms separately, which naturally provides an upper bound for $\text{\emph{var}}(T)$. Let $\sigma^2_Z = \text{\emph{var}}[\widetilde{T}]$. We shall provide explicit expressions of $\sigma^2_Z$ later. Then, 
$\E(U_{R,k}^2 \mid R = r) = \prod_{j=1}^k \E(1 - \nu \widetilde{T}_j)^2$ due to independence. Additionally, $\E(1 - \nu \widetilde{T}_j)^2 = (1 - \nu \mu)^2 + \nu^2 \sigma^2_Z$. We have the following result.
\begin{theorem}\label{thm:variance}
Define $\alpha = |1 - \nu \mu| < 1$ and $\beta^2 = \alpha^2 + \nu^2 \sigma^2_Z$. Let $\alpha < 1/(2e)$, $\beta< 1/(4e)$ and $R\sim \text{Geometric}(p)$, with $p < 1 - 4\beta^2e^2$. Then,
  \begin{align*}
  &\text{var}[\E(T\mid R)] \leq \dfrac{1}{(1 - 2e\alpha)^2} \dfrac{4\alpha^2 e^2 p}{1 - p - 4\alpha^2e^2},\\
  & \E[\text{var}(T \mid R)] \leq \dfrac{1 + 4e\beta}{1 - 4e\beta} \dfrac{1 - p}{1 - p - 4e^2\beta^2}.
  \end{align*}
  Consequently $\text{var}(T) < \infty$.
\end{theorem}
If the condition $\alpha < 1/(2e)$ is violated, then the conditional variance $\text{\emph{var}}[\E(T \mid R)]$ does not exist. Although it might seem that the more stringent assumption is $\beta<1/(4e)$ which involves the variance of $\widetilde{T}$, we emphasize here that this is achieved by increasing $N$.

We now turn our attention to $\sigma^2_Z$. For this, we shall make specific use of the fact that models under our consideration belong to the PEGM class. In particular, we study the random variable {\color{red} $$W \coloneqq f(Y;\theta')/f(Y; \theta),$$ }where $Y \sim p(\cdot;\theta)$ and $\theta, \theta' \in \Theta$. 
\begin{proposition}\label{prop:var_W}
    When $p(\cdot;\theta)$ is a PEGM and $2\theta' - \theta \in \Theta$, then: 
    $$\text{var}(W) = \dfrac{z(2\theta' - \theta)}{z(\theta)} - \dfrac{z^2(\theta')}{z^2(\theta)}.$$
\end{proposition}
This immediately implies that $\sigma^2_Z = N^{-1} [z(2\theta - \phi)/z(\phi) - z^2(\theta)/z^2(\phi)]=\mathcal{O}(N^{-1})$.

\subsection{Choosing $\nu$}
\label{sec:choosing_nu}
Crucially, the choice of $\nu$ controls both the numerical stability and the Monte Carlo efficiency of $T$. The infinite series $g_{\nu}(\mu)$ is effectively a Taylor expansion about 0. Therefore, both the truncation error and the variance improve as $|1-\nu\mu|$ shrinks. In practice, we run a pilot simulation to obtain $M$ independent replicates of $\widetilde{T}(\theta)$ to obtain $\hat{\mu}_{pilot} = M^{-1} \sum_{m=1}^M \widetilde{T}(\theta)$, and set $\nu = \alpha/\hat{\mu}_{pilot}$, where $\alpha \in (0, 2)$, so that $\nu\mu \approx \alpha$. Taking $\alpha = 1$ targets $\nu\mu \approx 1$, and choosing $\alpha < 1$ adds a conservative buffer to keep $|1-\nu\mu| < 1$ with high probability, ensuring convergence of $g_{\nu}(\mu)$ even when $\hat{\mu}_{pilot}$ is noisy. {\color{red} Additional implementational details are provided in Section \ref{sec:numerical_results}. }

\subsection{The Pseudo-marginal Sampler}
The proposed estimator can be used to conduct a valid pseudo-marginal algorithm. We now discuss specific details. Suppose $q(\cdot \mid \theta)$ is the proposal distribution. Then to make a Metropolis-Hastings move, we need to compute $\alpha_{MH}(\theta, \theta')$, which is given by:
$$\min\left\lbrace \dfrac{\prod_{l=1}^n f(\x_l; \theta') \pi(\theta') q(\theta \mid \theta') [z(\theta)]^n }{\prod_{l=1}^n f(\x_l;\theta) \pi(\theta) q(\theta' \mid \theta)[z(\theta')]^n},\; 1 \right\rbrace.$$
A valid pseudo-marginal algorithm will replace the intractable $[z(\theta)]^{-n}$ in the likelihood by its unbiased estimate. Also, for a suitably chosen tuning parameter $\nu$, let $T = T(\theta)$ be the unbiased estimator of $1/[\nu z(\theta)/z(\phi)]^n$ defined previously. Algorithm \ref{algo:algo} details the updates from step $t$ to step $t+1$.
\begin{algorithm}[!b]
    \caption{Pseudo-marginal sampler}
    \label{algo:algo}
    {\bf Input}: $\theta_t$ [current state], $\theta' \sim q(\cdot \mid \theta)$ [proposal], $\mathcal{D}$ [data], $N$ [number of Monte Carlo samples], $\nu_t$ [current tuning parameter], $T(\theta_t)$ [unbiased estimator of $1/[\nu z(\theta_t)/z(\phi_t)]^n$]

    {\bf Output}: $\theta_{t+1}$, $T(\theta_{t+1})$
    \begin{algorithmic}
\State 1. Set $\phi' = \text{diag}(\theta')$,  compute $\nu'$, construct $T(\theta')$.
\State 2. Compute
\begin{align}
\label{eq:alpha_IND}
&\alpha_{IND} = \alpha_{IND}\{(\theta_t, T(\theta_t)); (\theta', T(\theta'))\} \nonumber \\
& = \dfrac{\prod_{l=1}^n f(\x_l; \theta') \pi(\theta') q(\theta_t \mid \theta') [\nu'/z(\phi')]^n T(\theta')}{\prod_{l=1}^n f(\x_l; \theta_t) \pi(\theta_t) q(\theta' \mid \theta_t) [\nu_t/z(\phi_t)]^n T(\theta_t)}.
\end{align}

\If{{\color{red} $\widetilde{U} \sim \text{Uniform}(0,1) \leq  \min\{\alpha_{IND}, 1\}$}}
    \State $\theta_{t+1} = \theta'$, $\nu_{t+1} = \nu'$, $T(\theta_{t+1}) = T(\theta')$.
    \Else{\State $\theta_{t+1} = \theta_{t}$, $\nu_{t+1} = \nu_{t}$, $T(\theta_{t+1}) = T(\theta_{t})$.}
\EndIf
\end{algorithmic} 
\end{algorithm}

One issue with the sampler in Algorithm \ref{algo:algo} is that $T(\theta)$ is not almost surely non-negative. This is typical of randomized sum-estimators \citep{jacob2015nonnegative}. To deal with this, we define the non-negative posterior $|\hat{\pi}(\theta\mid \mathcal{D}, u)| \propto \prod_{l=1}^n f(\x;\theta) |T(\theta)| \pi(\theta)$, and run a pseudo-marginal chain with acceptance probability:
\small
\begin{align*}
 &\tilde{\alpha}_{PM}(\theta, \theta')=\min \left\lbrace \dfrac{|\hat{\pi}(\theta' \mid \mathcal{D}, u')| p(u') q(\theta \mid \theta') p(u)}{|\hat{\pi}(\theta \mid \mathcal{D}, u)| p(u) q(\theta' \mid \theta) p(u')},\; 1 \right\rbrace,
\end{align*} 
\normalsize
and keep track of $\sigma(\theta) \coloneqq \text{sgn}(T(\theta))$. {\color{red} Here, $u$ denotes all auxiliary random variables required for the unbiased estimation of the likelihood. This includes $R$ and $T$.} Finally, expectations with respect to the true posterior can be recovered by reweighting with the signs. Indeed, for any function $h(\theta)$, 
\begin{align*}
    \E_{\pi(\theta\mid \mathcal{D})}[h(\theta)] & = \int_{\theta, u} h(\theta) \pi(\theta, u\mid \mathcal{D}) d\theta du\\
    & = \dfrac{\int_{\theta, u} h(\theta) \sigma(\theta) |\hat{\pi}(\theta, u\mid \mathcal{D})| d\theta du}{\int_{\theta, u} \sigma(\theta) |\hat{\pi}(\theta, u\mid \mathcal{D})| d\theta du},
\end{align*}
since $\sigma(\theta)|\hat{\pi}(\theta, u \mid \mathcal{D})| = \hat{\pi}(\theta \mid \mathcal{D},u)$; see also \citet{lyne2015russian}. 

While Algorithm \ref{algo:algo} is an exact approximation of the true target $\pi(\theta\mid \x)$, it comes at an additional computational cost. In particular, for choosing the tuning parameter $\nu$ carefully to maintain finite variance of $T$, pilot estimates need to be constructed within each MCMC iteration. This becomes prohibitive when a large number of MCMC iterations is used. Additionally, ergodicity properties of the chain are not guaranteed even when the true chain, i.e. an M--H chain with $\alpha_{MH}$ as the acceptance probability, is ergodic \citep[Theorem 8]{andrieu2009pseudo}. To address these issues, in the next section, we also consider a \emph{noisy} alternative sampler.

\section{\uppercase{The noisy sampler}}
{\color{red} Although the pseudo-marginal sampler developed in the previous section targets the correct posterior, constructing the unbiased estimator $T$ at every MCMC iteration can become expensive as the dimension grows. Indeed, evaluating $T$ requires computing $R(R+1)/2$ copies of $\tilde{T}$, each of which needs $O(p)$ samples from $p(\cdot;\phi)$. This motivates developing a computationally cheaper but \emph{noisy sampler} targeting the posterior that no longer estimates the M--H ratio unbiasedly. }

For the noisy sampler, we target estimating $\alpha_{MH}$ in the log scale. Recall the unbiased estimator $\widetilde{T}(\theta)$ of $z(\theta)/z(\phi)$. In fact, $\widetilde{T}(\theta)$ is almost surely consistent. Additionally, if the support of $\x$ is bounded, then approximating $\alpha_{MH}$ in the log-scale is natural. This motivates the following estimate of $\log R_{MH}(\theta, \theta')$:
\begin{align}
    V(\theta, \theta') &=  \sum_{l=1}^n \log \dfrac{f(\x_l; \theta')}{f(\x_l;\theta)} + \log \dfrac{q(\theta\mid \theta') \pi(\theta')}{q(\theta'\mid \theta)\pi(\theta)} \nonumber\\
    &+ n \log \dfrac{\widetilde{T}(\theta)}{\widetilde{T}(\theta')} - n \log \dfrac{z(\phi')}{z(\phi)}.
\end{align}
The resulting noisy MCMC algorithm is given in Algorithm \ref{algo:algo_noisy}. Naturally, $\pi(\theta \mid \mathcal{D})$ is not the invariant distribution of this chain. However, one can expect that as $N$ increases, the approximation quality should improve. Moreover, one can ask whether the approximating chain inherits ergodicity properties of the original chain that uses $\alpha_{MH}$. We study this next formally.
\begin{algorithm}[!b]
    \caption{Noisy sampler}
    \label{algo:algo_noisy}
    {\bf Input}: $\theta_t$ [current state], $\theta' \sim q(\cdot \mid \theta)$ [proposal], $\mathcal{D}$ [data], $N$ [number of Monte Carlo samples]

    {\bf Output}: $\theta_{t+1}$
    \begin{algorithmic}
\State 1. Set $\phi = \text{diag}(\theta)$, $\phi' = \text{diag}(\theta')$, construct $\widetilde{T}(\theta)$, $\widetilde{T}(\theta')$.
\State 2. Compute $V(\theta, \theta')$.
\State {\color{red} 3. Sample $\widetilde{U} \sim  \text{Uniform}(0,1)$.}
\If{{\color{red} $\log \widetilde{U} \leq  V(\theta, \theta')$}}
    \State $\theta_{t+1} = \theta'$.
    \Else {\State $\theta_{t+1}=\theta_t$.}
\EndIf
\end{algorithmic} 
\end{algorithm}

Let $P(\theta, \cdot)$ and $\hat{P}_N(\theta, \cdot)$ be the transition kernels resulting from the acceptance probabilities $\min\{R_{MH}(\theta, \theta'),1\}$ and $\min\{e^{V(\theta, \theta')}, 1\}$. A Markov chain with initial value $\theta_0\in \Theta$, transition kernel $P$ and invariant distribution $\pi(\cdot\mid \mathcal{D})$ is said to be uniformly ergodic if $\norm{\delta_{\theta_0} P^t - \pi(\cdot \mid \mathcal{D})}_{TV} \leq C \rho^t$ for some $0<C< \infty$ and $\rho< 1$. Here $P^t$ is the $t$-th step transition kernel induced by $P$ {\color{red} and $\delta_{\theta_0} P^t$ is the distribution of the chain at the $t$-th step with $\theta_0$ as the initial value.} Suppose we run the approximate chain $\hat{P}_N$ with initial value $\theta_0$. Then we have the following result.
\begin{theorem}\label{thm:noisy_sampler}
    Suppose $\Theta = \{\theta \in \mathbb{R}^{p \times p}: \theta_{jk} = \theta_{kj}, |\theta_{jk}|\leq B, j, k = 1, \ldots, p\}$ for some $B>0$. The random variable $Y \sim p(\cdot;\theta)$ has bounded support. Let the prior $\pi(\theta)$ and the proposal $q(\cdot \mid \theta)$ be continuous for every $\theta \in \Theta$. Then:
    \begin{enumerate}
    \item  $P$ is uniformly ergodic in $\pi(\cdot \mid \mathcal{D})$ for every initial value $\theta_0 \in \Theta$ with some $C>0$, and some $\rho < 1$.
    \item Additionally, 
    $$\sup_{\theta_0} \norm{P^t - \hat{P}_N^t} \leq K/\sqrt{N}, $$
    where $K$ depends on $\pi(\theta)$, $q(\cdot \mid \theta)$ and $B$, $\rho$.
    \end{enumerate}
\end{theorem}
As a direct consequence of Theorem \ref{thm:noisy_sampler}, we get the following corollary.
\begin{corollary}\label{cor:cor1}
    Under conditions of Theorem \ref{thm:noisy_sampler}, $\hat{P}_N$ is also uniformly ergodic as $N \to \infty$.
\end{corollary}
The bounded support assumption of $Y \sim p(\cdot;\theta)$ is critical for approximating the M--H ratio in log-scale. Notably, many popular PEGMs satisfy this criterion, e.g. the Ising model, truncated Poisson graphical model etc.
Theorem \ref{thm:noisy_sampler} is similar to Theorem 3.2 of \cite{alquier2016noisy} but there the authors implicitly assume that sampling $\x \sim p(\cdot;\theta)$ is possible. This is true for perfect samplers but in practice Gibbs samplers are generally used due to the convenient univariate conditional distributions and lack of scalability of perfect samplers in high dimensions. In contrast, our approach does not presuppose the existence of perfect samplers, and has the benefit that no inner Gibbs chain is needed to implement it.

\section{\uppercase{Numerical Experiments}}\label{sec:numerical_results}

\subsection{Calibration of $\nu$}
As discussed in Section \ref{sec:choosing_nu}, the choice of $\nu$ controls the quality of $T$. One condition to ensure that $T$ is well-behaved is that $|1 - \nu \mu|<1$ where $\mu$ is estimated by pilot runs of $\tilde{T}$. In fact, it is only a sufficient condition. For faster convergence, we want it to be close to 0. Recall, we set $\nu = \alpha/\hat{\mu}_{pilot}$ where $\alpha \in (0,2)$. Here, we assess the sensitivity of $|1 - \nu\mu|$ to the choice of $\alpha$ and the importance sample size $N$ across varying dimensions $p$. Our experiments are done for the Ising model.

Fixing the number of pilot replicates at $M = 100$, Table~\ref{tab:nu_calibration} reports the average value of $|1 - \nu\hat{\mu}_{pilot}|$ across 100 replications for varying dimensions $p$ and importance sample sizes $N$. When $\alpha = 1$, the quantity $|1 - \nu\hat{\mu}_{pilot}|$ decreases steadily as $N$ increases, approaching zero for large $N$ across all dimensions considered. 
However, the rate of convergence slows with increasing $p$, requiring substantially larger importance samples to achieve a small value of $|1 - \nu\hat{\mu}_{pilot}|$ in higher dimensions. Setting $\alpha = 0.5$ yields values that stabilize near $0.5$ across all $N$, even for $p = 50$ and $100$, as expected since $\nu\hat{\mu}_{pilot} \approx 0.5$ by construction.

Based on these results, we recommend $\alpha = 1$, for moderate $p$ as it yields $|1 - \nu\mu|$ closest to zero. In higher-dimensions, where large $N$ may be computationally
prohibitive, setting $\alpha < 1$ provides a reliable safeguard by ensuring $|1 - \nu\mu| < 1$ regardless of the accuracy of the pilot estimate.

\begin{table}[!t]
    \centering
    \caption{Average $|1 - \nu\mu|$, where $\nu = \frac{\alpha}{\hat{\mu}_{\text{pilot}}}$ across 100 replications. Theory requires a value less than $1$. Closer to $0$ is better.}
    \label{tab:nu_calibration}
    \scalebox{0.8}{
    \begin{tabular}{l cccccc}
        \toprule
        \multicolumn{7}{c}
        {\textit{$\alpha = 1$}} \\
         \addlinespace
         \midrule
        N & 1000 & 5000 & 10000 & 50000 & 100000 & 500000 \\
        \midrule
        $p = 5$   & 0.01 & 0.01 & 0.00 & 0.00 & 0.00 & 0.00 \\
        $p = 50$  & 0.22 & 0.08 & 0.07 & 0.03 & 0.02 & 0.01 \\
        $p = 100$ & 1.94 & 0.72 & 0.52 & 0.16 & 0.13 & 0.06 \\
        \midrule
        \addlinespace
        \multicolumn{7}{c}
        {\textit{$\alpha = 0.5$}} \\
        \addlinespace
        \midrule
         N & 1000 & 5000 & 10000 & 50000 & 100000 & 500000 \\
        \midrule
        $p = 5$   & 0.10 & 0.10 & 0.10 & 0.10 & 0.10 & 0.10 \\
        $p = 50$  & 0.48 & 0.50 & 0.50 & 0.51 & 0.50 & 0.50 \\
        $p = 100$ & 0.90 & 0.42 & 0.40 & 0.48 & 0.48 & 0.49 \\
        \bottomrule
    \end{tabular}
    }
\end{table}

\subsection{Comparison of the proposed method with alternatives}
We compare the performance of the proposed exact-pseudo-marginal (PM) sampler and the noisy (N) sampler with the exchange algorithm (EX) in low and high-dimensional Ising models ($p = 3, 5, 20, 50, 70, 100$). The auxiliary variable in the exchange algorithm is drawn using an inner Gibbs sampler. For all the samplers, we consider two proposal distributions: the symmetric random walk $q(\theta' \mid \theta) \coloneqq \mathrm{N}(\theta, \sigma^2\mathrm{I}_{p(p+1)/2})$ (RW) and the approximate Langevin (L) proposal $q(\theta'\mid \theta) \sim \mathrm{N}(\theta + \gamma \hat{\nabla} \log \pi(\theta\mid \mathcal{D}), \sigma^2\mathrm{I}_{p(p+1)/2} )$ with $\sigma^2$ and $\gamma>0$ being the step-sizes and $\hat{\nabla} \log \pi(\theta \mid \mathcal{D})$ is some estimate of the true gradient of the log-posterior. Specific details of construction of such proposals are given in Supplementary Section~\ref{sec:approx_gradient}. The prior $\pi(\theta)$ for all the cases is a product Laplace distribution, i.e. $\pi(\theta \mid \lambda) = \prod_{j\leq k} \pi(\theta_{jk}\mid \lambda)$ where $-\log\pi(\theta_{jk}\mid \lambda) = \lambda |\theta_{jk}| + C$ for $\lambda>0$. The hyperparameter $\lambda$ is chosen to maximize the out-of-sample log-likelihood on a test set. 

For $p = 3, 5$, we consider $n = 100$ observations generated by a \emph{dense} true parameter $\theta_0$ with $\theta_{0,jk}= -1 $ with probability $0.9$, and zero with probability $0.1$. For high-dimensional settings ($p\geq 20$), we set $n=200$ and a \emph{sparse parameter}: $\theta_{0, jk} = -3$ with probability $0.02$, and zero with remaining probability.  The number of Monte Carlo samples $N$ is $5,000$ for $p = 3, 5, 20$; it is $10,000$ for $p=50$ and $N =50,000$ for $p=70$, $ p= 100$. We evaluate the methods in three aspects: (1) runtime, (2) mixing, i.e., the samplers' ability to move into high-posterior regions quickly, measured via the effective sample sizes computed as: $T_0/(1 + \sum_{k=0}^\infty \rho_k)$ where $\rho_k$ is the $k$-lag autocorrelation of the chain and $T_0$ is the total number of MCMC samples, and (3) their ability to recover the true parameter $\theta_0$ which is measured by $||\theta - \hat{\theta}||_F^2/p^2$ where $\hat{\theta}$ is the posterior mean of each of these samplers, and the scaling by the total number of parameters $p^2$ ensures the results are comparable across $p$. All methods were implemented in \texttt{Rcpp} on a single Dell HPC node (dual 64-core AMD EPYC “Milan,” 256 GB RAM, 100 Gbps HDR InfiniBand) and each run is restricted to 25 CPU cores. We use {\color{red} a total of }$5,000$ MCMC iterations with {\color{red} the initial} $2,000$ samples discarded as burn-in.

\begin{table}[!t]

\centering
\caption{ Average runtime (minutes)  across 30 data sets. ``--" indicates omitted runs due to poor mixing for RW.} 
\label{tab:runtime_N5000}
\scalebox{0.8}{ 
\begin{tabular}{ccccccc}
\hline
Sampler &  \multicolumn{2}{c}{PM} & \multicolumn{2}{c}{N} & \multicolumn{2}{c}{EX} \\
\hline
Proposal & RW & L & RW & L & RW & L  \\
\hline
$ p = 3$ & 0.873 & 0.861 & 0.037 & 0.093 & 0.001 & 0.096 \\
$p = 5$ & 1.589 & 1.533 & 0.068 & 0.175 & 0.002 & 0.192 \\
$p = 20$ & 10.427 & 3.500 & 0.219 & 0.688 & 0.014 & 0.750\\
$p = 50$ & 57.646 & 48.848 & 2.367  & 4.827 & 0.041 & 5.441 \\
$p=70$ & -- & 109.614 & -- & 45.316 & -- & 46.563 \\
$p=100$ & -- & 344.388 & -- & 88.869 & -- & 82.395 \\
\hline
\end{tabular}
}
\end{table}

\begin{table}[!b]
\centering
\caption{Average Mean Effective Sample Size  across 30 data sets. ``--" indicates omitted runs due to poor mixing for RW.}
\label{tab:ESS}
\scalebox{0.8}{
\begin{tabular}{ccccccc}
\hline
Sampler &  \multicolumn{2}{c}{PM} & \multicolumn{2}{c}{N} & \multicolumn{2}{c}{EX} \\
\hline
Proposal & RW & L & RW & L & RW  & L \\
\hline
$ p = 3$ &  88.7 & 86.8 & 85.9 & 86.5 & 79.5 & 80.5\\
$p = 5$ &  130.3 & 128.4 & 81.1 &  84.6 & 79.4 & 83.4 \\
$p = 20$ & 150.3 & 176.3 & 79.0 & 79.2 & 79.0 & 80.0 \\
$p = 50$ & 381.8 & 368.3 & 78.8 & 78.9 & 78.6 & 79.0 \\
$p=70$ & -- & 661.3 & -- & 79.3 & -- & 79.6 \\ 
$p=100$ &  -- & 1052.0 & -- & 78.5 & -- & 79.6\\
\hline
\end{tabular}
}
\end{table}

\begin{figure}[!b]
\vspace{-0.5cm}
    \centering
    \includegraphics[height = 7.5cm, width = 7.5cm]{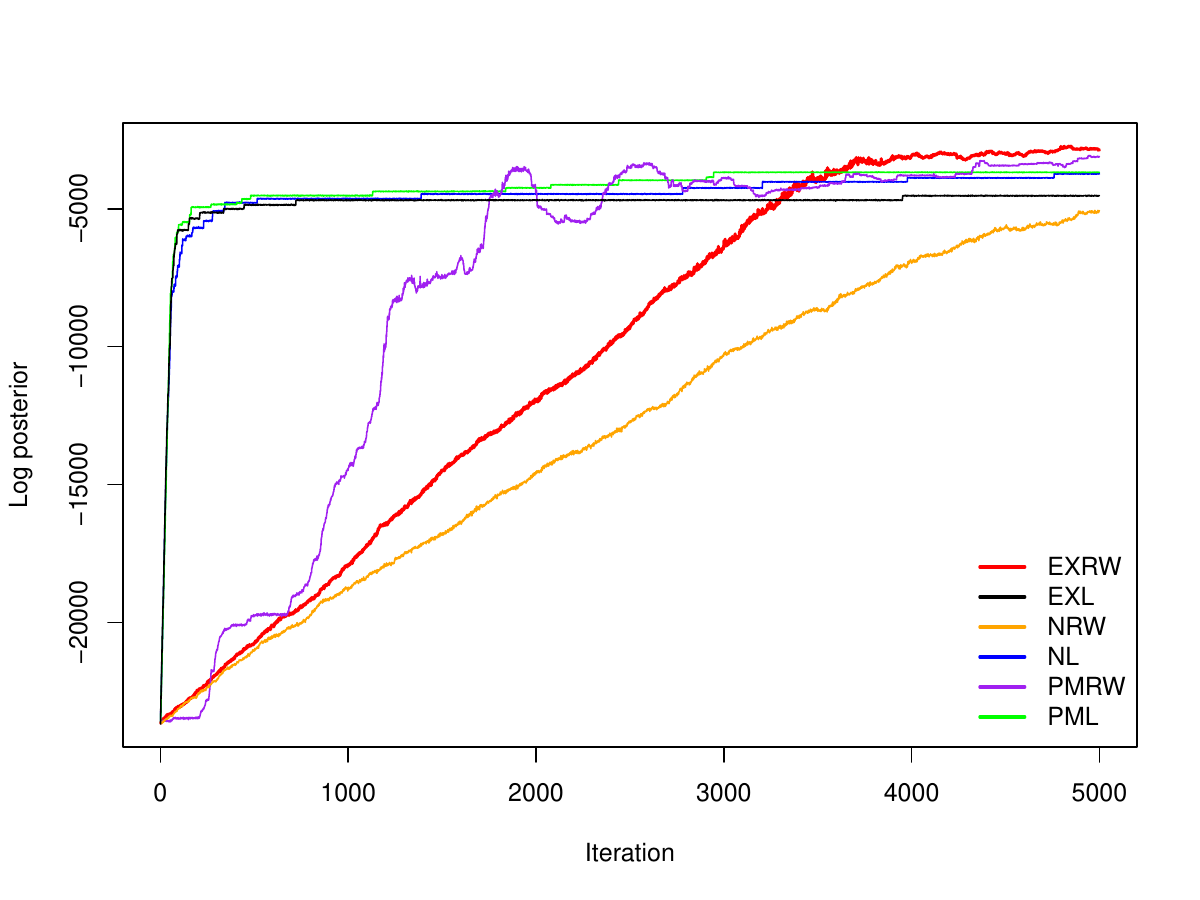}
    \caption{Log posterior trace plots for $p = 20$. 
    }
    \label{fig:log_pos}
\end{figure}

\begin{table*}
\centering
\caption{Mean (standard deviation) of MSE = $||\hat{\theta} - \theta_0||_F^2/p^2$, across 30 data sets. ``--" indicates omitted runs due to poor mixing for RW.
} 
\scalebox{0.8}{ 
\begin{tabular}{cccccccc}
\hline
Sampler &  \multicolumn{2}{c}{PM} & \multicolumn{2}{c}{N} & \multicolumn{2}{c}{EX} \\
\hline
Proposal & RW & L & RW & L & RW & L \\
\hline
$ p = 3$ & 0.084 (0.015) & 0.088 (0.024) & 0.091 (0.027) & 0.103 (0.036) & 0.101 (0.028) & 0.108 (0.033) \\
$p = 5$ & 0.077 (0.012) & 0.067 (0.017)  & 0.079 (0.011) & 0.077 (0.016) & 0.087 (0.011)  & 0.108 (0.029)\\
$p = 20$ & 0.022 (0.003) & 0.019 (0.0004) & 0.034 (0.002)& 0.019 (0.0004) & 0.022 (0.002) & 0.019 (0.0004) \\
$p = 50$ & 0.018 (0.0011) & 0.009 (0.0001) & 0.022 (0.0002) & 0.009 (0.0001) & 0.016 (0.0002) & 0.009 (0.0001) \\
$p=70$ & -- & 0.007(0.00006)& -- &0.007(0.00005) & -- & 0.007 (0.00004)\\
$p = 100$ & -- & 0.005 (0.00005)
& -- & 0.005 (0.00002) & -- & 0.005 (0.00003) \\
\hline
\end{tabular}
}
\label{tab:norm_N5000}
\end{table*}

\begin{figure*}[!h]
\begin{subfigure}{0.05\textwidth}
    \adjustbox{raise = 10ex}{\includegraphics[height = 2cm, width = 1cm]{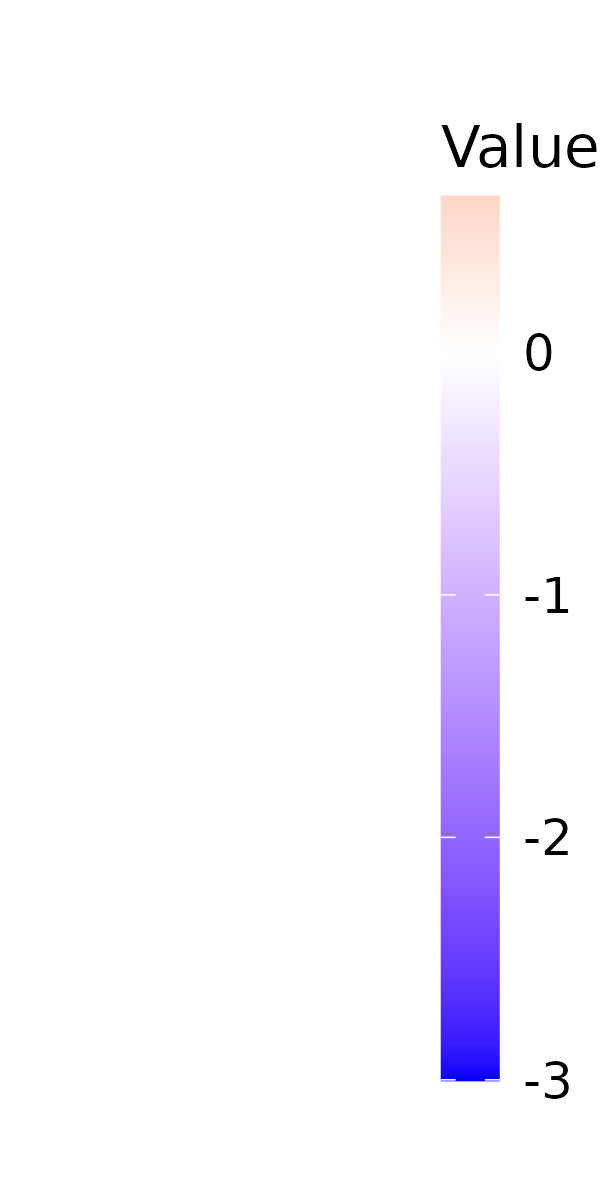}}
\end{subfigure}
\begin{subfigure}{0.3\textwidth}
    \centering
    \includegraphics[height = 3.5cm, width = 3.5cm]{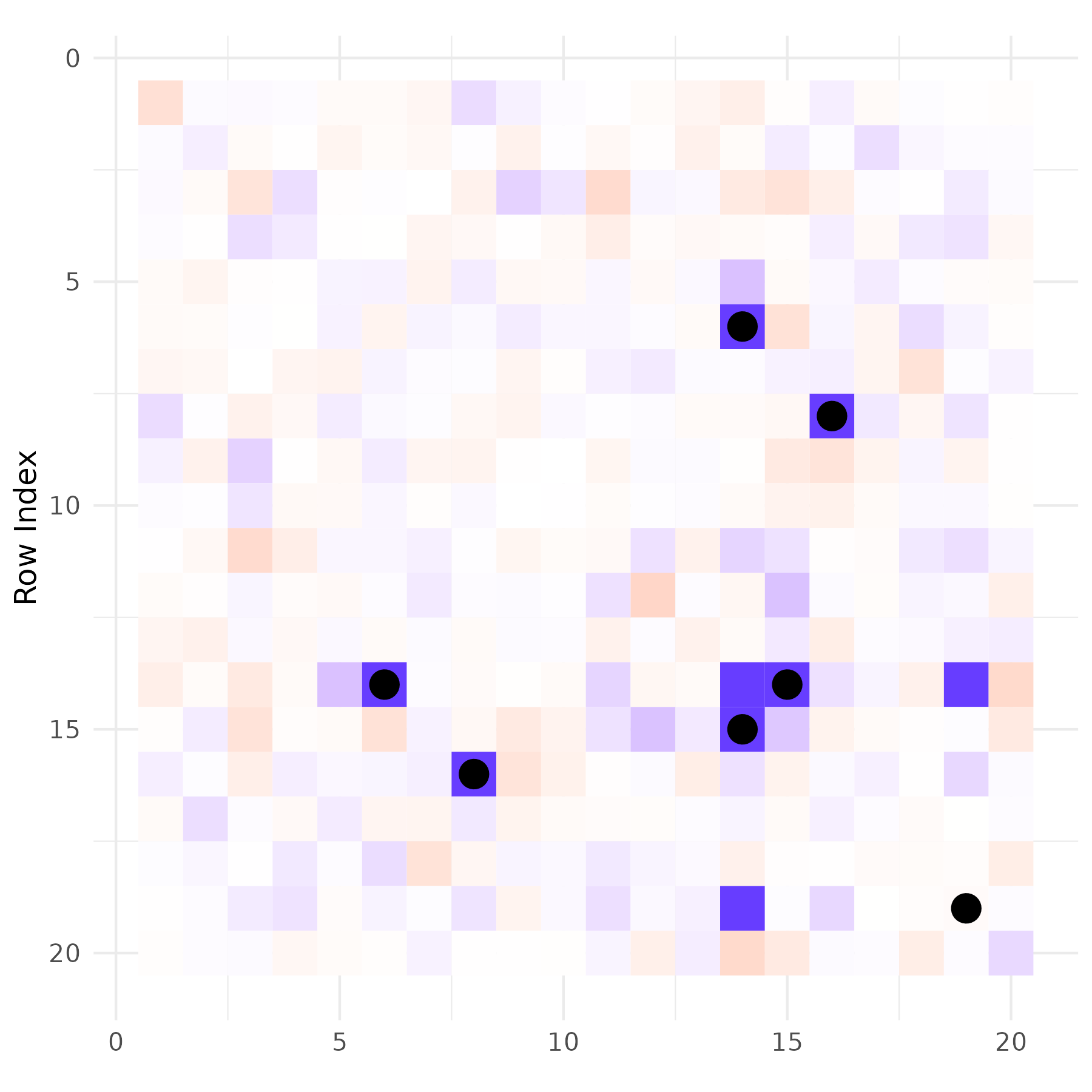}
    \caption{}
\end{subfigure}%
\begin{subfigure}{0.3\textwidth}
    \centering
    \includegraphics[height = 3.5cm, width = 3.5cm]{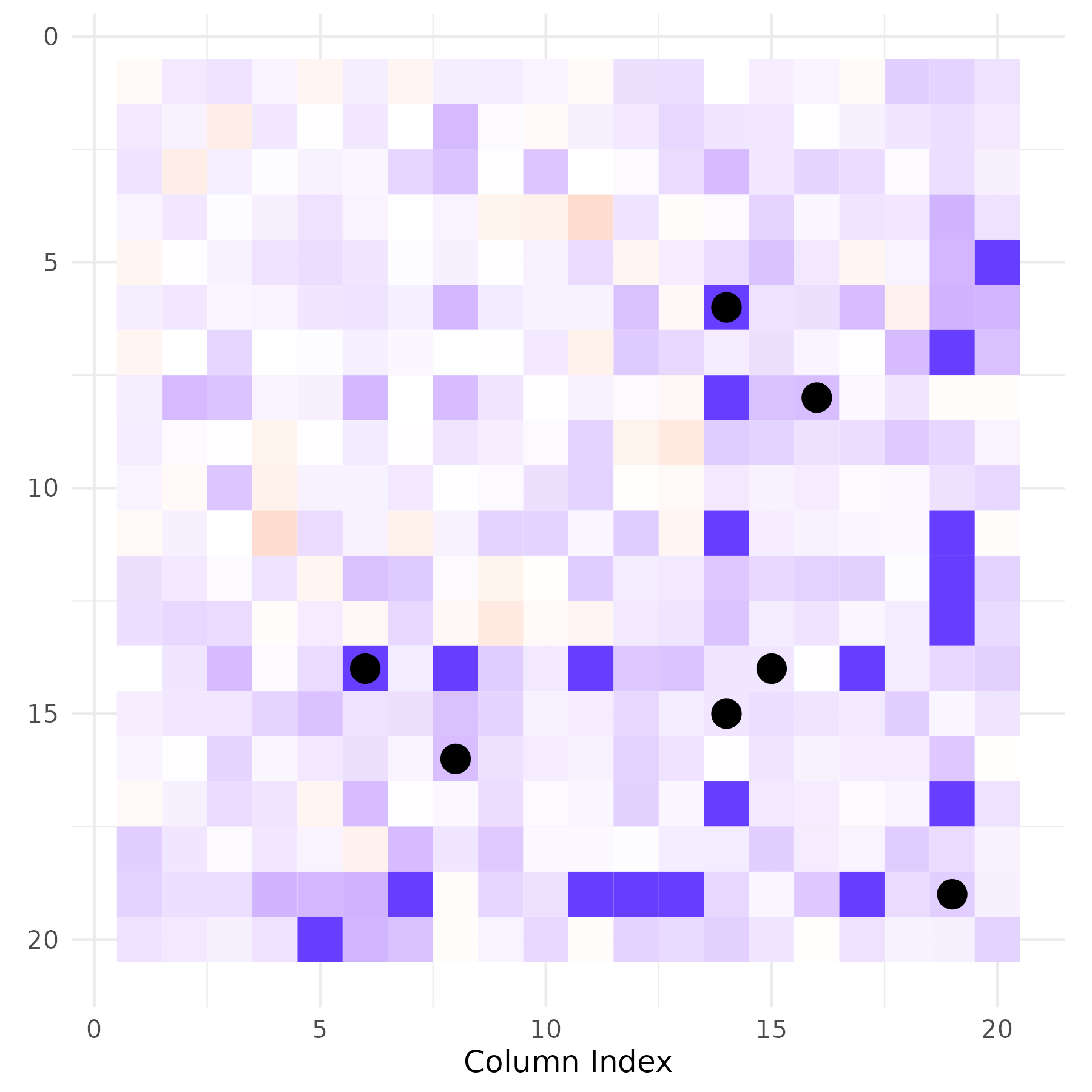}
    \caption{}
\end{subfigure} 
\begin{subfigure}{0.3\textwidth}
    \centering
    \includegraphics[height = 3.5cm, width = 3.5cm]{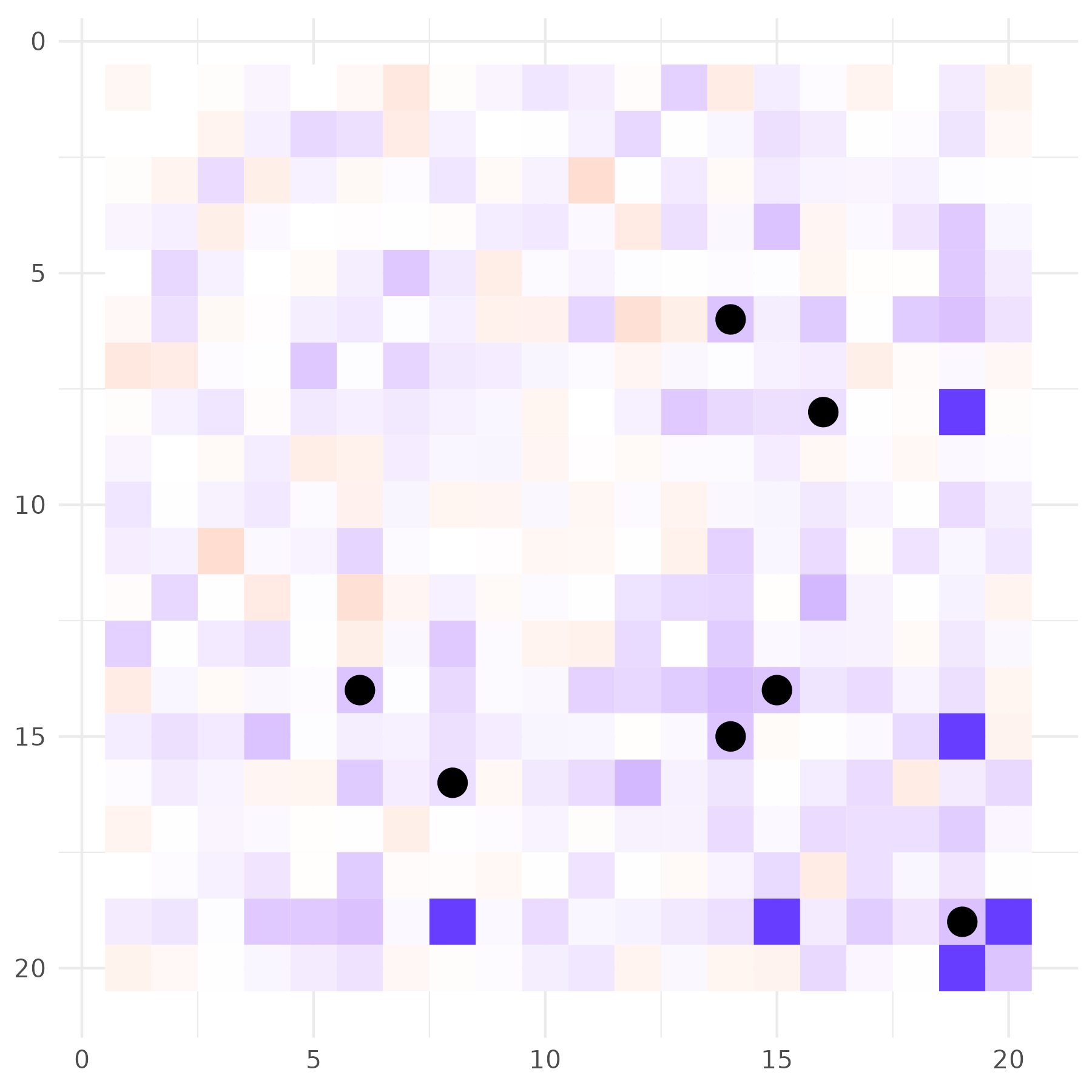}
    \caption{}
\end{subfigure}\\
\caption{Posterior mean estimates of the parameter matrix $\theta$ for $p = 20$. (a) PMRW (b) NRW (c) EXRW. The true non-zero elements in $\theta_0$ have value $-3$, with their locations indicated by black dots.\label{fig:heatmap} 
} 
\end{figure*}

Table \ref{tab:runtime_N5000} shows the pseudo-marginal sampler has the maximum runtime while the noisy version has comparable runtime to the exchange algorithm. We emphasize here that the exchange algorithm is not implemented with a perfect sampler. We expect the runtime of the exchange algorithm to significantly increase if that were the case. In fact, as the dimension grows, runtime of the noisy sampler in Algorithm \ref{algo:algo_noisy} and the exchange algorithm become almost the same. More importantly, Table \ref{tab:ESS} shows that effective sample sizes from the pseudo-marginal chain are far better than the other two samplers. Indeed, when PM(L) and EX(L) are compared in terms of ESS/minute, at $p = 70$, these numbers are 6.03 and 1.70, respectively. At $p = 100$, they are 3.05 and 0.96 for PM(L) and EX(L). In Table \ref{tab:norm_N5000}, we report $||\hat{\theta} - \theta_0||_F^2/p^2$. All samplers perform comparably in terms of recovering the true parameter.  In summary, our findings suggest that in low-dimensions, with moderate computational budget, pseudo-marginal sampler in Algorithm \ref{algo:algo} is preferable over the other two choices, whereas in high-dimensions the noisy sampler in Algorithm \ref{algo:algo_noisy} performs better. For all these samplers, Figure \ref{fig:log_pos} shows that the proposed (approximate) gradient-based proposals move to high-posterior regions much faster than simple random-walk proposals. Figure~\ref{fig:heatmap} shows a heatmap of  posterior mean estimates for $p=20$ under different methods. {\color{red} See Supplementary Section~\ref{sec:supp_num} for additional results.}

\section{\uppercase{Data Applications}}\label{sec:real_data}
We demonstrate the proposed method using the $MovieLens \ 32M$ dataset, which contains 32 million movie ratings provided by $200,948$ users across $87,585$ films, with ratings ranging from 0 to 5 in increments of 0.5 (\href{https://grouplens.org/datasets/movielens/}{https://grouplens.org/datasets/movielens/}). 
We select $p = 50$ most popular movies that were rated by the same group of $n = 448$ viewers. To dichotomize the ratings, we code movies with ratings of $5$ as $1$, whereas ratings of $4.5$ and below are coded as $0$. Let $X_{ij}$ denote the preference of user $i$ for movie $j$. We consider fitting an Ising model to this data by assuming $X_{i} \overset{i.i.d.}{\sim} \text{Ising}(\theta)$, for $i=1,\ldots,n$, and $\theta\in\mathbb{R}^{p\times p}$. A positive estimated value of $\theta_{jk}$ can now be interpreted as a common preference for movies $j$ and $k$ across users, whereas a negative value would indicate opposite preferences. We use the product Laplace prior $\pi(\theta \mid \lambda = 40 )$ using the samplers PM(L), N(L), EX(L). We set $N =70,000$ and run $10,000$ MCMC iterations with 7000 burn-in samples. To assess consistency across methods, we compare the signs of the posterior mean estimates $\hat{\theta}_{jk}$, after thresholding at $|\hat{\theta}_{jk}| > 0.1$. For each pair of methods, we calculate the proportion of entries in $\theta$ where the two methods agree in sign (both positive, both negative, or both zero). 
The pseudo-marginal method shows the highest agreement with the noisy method ($62\%$), while both show weaker agreement with the exchange method ($54\%$ and $56\%$ for PM(L) and N(L) respectively).
Further analysis is provided in Supplementary Section~\ref{sec:supp_data}.

\section{\uppercase{Conclusions}}
In this article, we propose two alternatives for posterior sampling in doubly-intractable models. The first one is an exact pseudo-marginal sampler that targets the correct posterior distribution, and the other is an approximate sampler. In particular, for the pseudo-marginal sampler, we develop an unbiased estimator of negative powers of the normalizing constant, and show that the resulting estimator has finite variance. For high-dimensional models, we also propose a noisy sampler, which inherits ergodicity properties of the original chain. Numerical experiments show that the pseudo-marginal chain has better mixing properties. The defining feature of our approach is that an \emph{inner loop} of a sequential sampler is not needed and both our proposals use the underlying \emph{independence model} for sampling purposes, which helps with scalability as well as mixing. This contrasts with existing alternatives such as the exchange algorithm, which presupposes a \emph{perfect sampler} \citep{propp1996exact}, but in practice, almost always uses an inner loop in a double MH type procedure \citep{liang2010double} in high dimensions.

Several future avenues of investigation could naturally build on the current work. Although we consider the Ising model, there is a large class of intractable graphical models that also consist of an underlying independence model, such as the Potts model \citep{potts1952some}, the Poisson graphical model \citep{besag1974spatial} and Boltzmann machines \citep{hinton2007boltzmann}. The proposed approach seems feasible in all these cases. Alternatives to Langevin, such as Hamiltonian Monte Carlo \citep{neal2011mcmc} could also be developed following our approach.

\section*{\uppercase{Code Availability}}
{\color{red}Code and usage examples are available at: \url{https://github.com/chenyujie1104/exact-approx-mcmc} }
\section*{\uppercase{Acknowledgments}}
{\color{red} 
Chakraborty and Bhadra acknowledge support from the US National Science Foundation Grant SES-2448704.}
\clearpage
\bibliographystyle{plainnat}
\bibliography{ref}

\clearpage
\section*{Checklist}



 \begin{enumerate}

 \item For all models and algorithms presented, check if you include:
 \begin{enumerate}
   \item A clear description of the mathematical setting, assumptions, algorithm, and/or model. [Yes, all models, algorithms and theoretical results are provided with clear list of assumptions.]
   \item An analysis of the properties and complexity (time, space, sample size) of any algorithm. [Yes, detailed runtime analysis of the proposed algorithms are provided and compared with existing ones, see Section \ref{sec:numerical_results}.]
   \item (Optional) Anonymized source code, with specification of all dependencies, including external libraries. [Yes]
 \end{enumerate}

 \item For any theoretical claim, check if you include:
 \begin{enumerate}
   \item Statements of the full set of assumptions of all theoretical results. [Yes, refer to Theorems \ref{thm:variance}, \ref{thm:noisy_sampler}.]
   \item Complete proofs of all theoretical results. [Yes, proofs are provided in the Supplement.]
   \item Clear explanations of any assumptions. [Yes.]     
 \end{enumerate}

 \item For all figures and tables that present empirical results, check if you include:
 \begin{enumerate}
   \item The code, data, and instructions needed to reproduce the main experimental results (either in the supplemental material or as a URL). [Yes.]
   \item All the training details (e.g., data splits, hyperparameters, how they were chosen). [Yes.]
         \item A clear definition of the specific measure or statistics and error bars (e.g., with respect to the random seed after running experiments multiple times). [Yes.]
         \item A description of the computing infrastructure used. (e.g., type of GPUs, internal cluster, or cloud provider). [Yes, see Section \ref{sec:numerical_results}.]
 \end{enumerate}

 \item If you are using existing assets (e.g., code, data, models) or curating/releasing new assets, check if you include:
 \begin{enumerate}
   \item Citations of the creator if your work uses existing assets. [Yes, see Section \ref{sec:real_data}.]
   \item The license information of the assets, if applicable. [Not Applicable]
   \item New assets either in the supplemental material or as a URL, if applicable. [Not Applicable]
   \item Information about consent from data providers/curators. [Not Applicable]
   \item Discussion of sensible content if applicable, e.g., personally identifiable information or offensive content. [Not Applicable]
 \end{enumerate}

 \item If you used crowdsourcing or conducted research with human subjects, check if you include:
 \begin{enumerate}
   \item The full text of instructions given to participants and screenshots. [Not Applicable]
   \item Descriptions of potential participant risks, with links to Institutional Review Board (IRB) approvals if applicable. [Not Applicable]
   \item The estimated hourly wage paid to participants and the total amount spent on participant compensation. [Not Applicable]
 \end{enumerate}

 \end{enumerate}

\clearpage
\onecolumn
\clearpage\pagebreak\newpage
\setcounter{secnumdepth}{3}
\setcounter{equation}{0}
\setcounter{page}{1}
\setcounter{table}{0}
\setcounter{section}{0}
\setcounter{subsection}{0}
\setcounter{figure}{0}
\setcounter{algorithm}{0}
\renewcommand{\theequation}{S.\arabic{equation}}
\renewcommand{\thesection}{S.\arabic{section}}
\renewcommand{\thepage}{S.\arabic{page}}
\renewcommand{\thetable}{S.\arabic{table}}
\renewcommand{\thefigure}{S.\arabic{figure}}
\renewcommand{\thealgorithm}{S.\arabic{algorithm}}

\section*{\uppercase{Supplementary Material}}

\section{Proofs}\label{sec:supp_proofs}
{\color{red} We first provide definitions of key quantities for the ease of readability. The main estimator in this work is: 
\begin{align*}
    T = \sum_{k=0}^R \dfrac{\gamma_k}{\P(R\geq k)} U_{R,k},
\end{align*}
where given $R=r$,
$${\color{red} U_{r,k} = \prod_{j=1}^k (1 - \nu \widetilde{T}_j), \quad 0< k \leq r.}$$
In the above display, 
\begin{equation*}
    \widetilde{T} = \widetilde{T}(\theta) = \frac{1}{N} \sum_{i=1}^N \frac{f(\y_i;\theta)}{f(\y_i;\phi)}, \quad \y_i \overset{iid}{\sim }p(\cdot;\phi).
\end{equation*}
}
\subsection{Proof of Proposition \ref{prop:finite_absolute_expectation}}
By construction, $\E(|U_{r,k}|) = m_k$ for all $r \geq k$. Next, $m_k = \prod_{j=1}^k \E(|V_j|) = \prod_{j=1}^k \E|1 - \nu \tilde{T}| < 1$, by assumption. Hence $a_k = m_k< \infty$. Furthermore, $m_{k+1} < m_k$. Thus, $\lim_{k \to \infty} \gamma_{k+1}a_{k+1}/\gamma_k a_k = \lim_{k \to \infty } [(n+k)/(k+1)] [a_{k+1}/a_k] < 1$. This implies that $\sum_{k=1}^\infty \gamma_k a_k < \infty$. 

\subsection{Proof of Proposition \ref{prop:var_W}}
Recall that $W = f(Y;\theta')/f(Y;\theta) = f(Y;\theta' -\theta)$, where $Y \sim p(\cdot;\theta)$. Also, $f^2(Y;\theta) = f(Y; 2\theta)$.
Hence,
\begin{align*}
    \E[W^2] = \E[f(Y; 2(\theta' -\theta))]& = \frac{1}{z(\theta)}\int_y f(y; 2(\theta' - \theta)) f(y;\theta) dy \\
    & = \frac{1}{z(\theta)}\int_y f(y; 2\theta' - \theta) dy = \dfrac{z(2\theta' - \theta)}{z(\theta)}.
\end{align*}
The proof follows by noticing that $\E[W] = z(\theta')/z(\theta)$.
\subsection{Proof of Theorem \ref{thm:variance}}
Following \cite{chopin2024towards}, we have:
\begin{align*}
    \text{\emph{var}}[\E(T\mid R)] =& \sum_{k=0}^\infty \gamma_k^2 (1 - \nu \mu)^{2k} \left[ \dfrac{1}{P(R\geq k)} - 1\right]\\
     &+ 2 \sum_{k=0}^\infty \sum_{l = k+1}^\infty \gamma_k \gamma_l (1 - \nu \mu)^{k+l} \left[ \dfrac{1}{P(R\geq k)} - 1\right] \\
      :=& A_1 + A_2.
\end{align*}
Next, we bound each of the terms $A_1$ and $A_2$ separately. For $A_1$, since $\gamma_k \leq [(n+k-1)e/(k-1)]^k$, we have:
\begin{align*}
    A_1 &\leq \sum_{k=0}^\infty (2\alpha e)^{2k} \left[ \dfrac{1}{P(R\geq k) } - 1\right]\\
    & = \sum_{k=0}^\infty \left[ \left(\dfrac{4\alpha^2e^2}{1 - p}\right)^{k} - (2\alpha e)^{2k}\right]\\
    & = \dfrac{4\alpha^2 e^2 p}{(1 - 4\alpha^2e^2)(1 - p - 4\alpha^2e^2)},
\end{align*}
where we used the assumption that $p < 1- 4e^2\beta^2 < 1 - 4e^2 \alpha^2$ since $\beta^2 = \alpha^2 + \nu^2 \sigma^2_Z$. 
We now consider $A_2$. We have:
\begin{align*}
    A_2=2 \sum_{k=0}^\infty \sum_{l = k+1}^\infty \gamma_k \gamma_l \alpha^{k+l} \left[ \dfrac{1}{P(R\geq k) } - 1\right]&\leq 2 \sum_{k=0}^\infty \gamma_k \alpha^k\left[ \dfrac{1}{P(R\geq k)} - 1\right] \sum_{l = k+1}^\infty (2e)^{l} \alpha^l \\
    &  \leq \frac{4e\alpha}{1 - 2e\alpha} \sum_{k=0}^\infty (2e\alpha)^{2k}\left[ \dfrac{1}{P(R\geq k) } - 1\right] \\
    & = \dfrac{4e\alpha}{1 - 2e\alpha}\dfrac{4\alpha^2 e^2 p}{(1 - 4\alpha^2e^2)(1 - p - 4\alpha^2e^2)}.
\end{align*}
This proves the first assertion. Now, we consider $\E[\text{\emph{var}}(T \mid R)]$. From the law of total variance, we have that $\E[\text{\emph{var}}(T \mid R)] \leq \E[\E(T^2 \mid R)]$. Next, recall $T \mid R = \sum_{k=0}^R \dfrac{\gamma_k}{\P(R\geq k)} U_{R,k}$. Hence,
\begin{align*}
    \E(T^2 \mid R) &= \sum_{k=0}^R \dfrac{\gamma_k^2}{[\P(R \geq k)]^2} \E(U_{R,k}^2)+ 2 \sum_{k=0}^{R-1} \sum_{l=k+1}^{R} \dfrac{\gamma_k \gamma_l}{\P(R \geq k) \P(R \geq l)} \E(U_{R,k}U_{R,l}).
\end{align*}
Since $U_{R,k} = \prod_{i=1}^k (1 - \hat{Z}_i)$ where $Z_i$ is independent of $Z_j$ for $i, j \leq k$, we obtain:
$$\E(U_{R,k}^2) = \prod_{i=1}^k \{(1 - \nu \mu)^2 + \nu^2 \sigma_Z^2\} = \beta^{2k}.$$
Moreover, by the Cauchy-Schwarz inequality, $\E(U_{R,k} U_{R, l}) \leq \E(|U_{R,k} U_{R,l}|) \leq [\E(U_{R,k}^2)]^{1/2} [\E(U_{R,l}^2)]^{1/2} = \beta^{k+l}$.
Thus,
\begin{align*}
    \E[T^2\mid R = r] \leq& \sum_{k=0}^R \dfrac{\gamma_k^2}{[\P(R \geq k)]^2} \beta^{2k} + 2 \sum_{k=0}^R \sum_{l=1}^{k+1} \dfrac{\gamma_k \gamma_l}{\P(R \geq k) \P(R \geq l)} \beta^{k+l}.
\end{align*}
This implies that for sufficiently large $r_0 \in \mathbb{N}$,
\begin{align*}
    &\sum_{r=0}^{r_0} \E[T^2 \mid R = r] \P[R=r] \leq  \sum_{r=0}^{r_0} \P(R = r) \sum_{k=0}^r \dfrac{\gamma_k^2 \beta^{2k}}{[\P(R \geq k)]^2}  + 2 \sum_{r=0}^{r_0} \P(R = r)\sum_{k=0}^{r-1}\sum_{l=k+1}^{r} \dfrac{\gamma_k \gamma_l \beta^{k+l}}{\P(R \geq k) \P(R \geq l)}.
\end{align*}
Now,
\begin{align*}
    \sum_{r=0}^{r_0} \P(R = r) \sum_{k=0}^r \dfrac{\gamma_k^2 \beta^{2k}}{[\P(R \geq k)]^2} &= \sum_{k=0}^{r_0} \dfrac{\gamma_k^2 \beta^{2k}}{[\P(R \geq k)]^2}  \sum_{r=k}^{r_0} \P(R = r)\\ &\leq \sum_{k=0}^r \dfrac{\gamma_k^2 \beta^{2k}}{[\P(R \geq k)]}\leq \sum_{k=0}^\infty \dfrac{\gamma_k^2 \beta^{2k}}{[\P(R \geq k)]} \\& \leq  \sum_{k=0}^\infty \dfrac{(2e\beta)^{2k}}{(1 -p)^k} = \dfrac{1 - p}{1 - p -4e^2\beta^2}.
\end{align*}
Similarly, 
\begin{align*}
    \sum_{r=0}^{r_0} \P(R = r)\sum_{k=0}^{r-1}\sum_{l=k+1}^{r} \dfrac{\gamma_k \gamma_l \beta^{k+l}}{\P(R \geq k) \P(R \geq l)}
 &= \sum_{k=0}^{r_0-1} \sum_{l=k+1}^{r_0} \dfrac{\gamma_k \gamma_l \beta^{k+l}}{\P(R \geq k) \P(R \geq l)} \sum_{r =l}^{r_0} \P(R = r) \\
    & \leq \sum_{k=0}^{r_0-1} \sum_{l=k+1}^{r_0} \dfrac{\gamma_k \gamma_l \beta^{k+l}}{\P(R \geq k)}\\
    &  \leq \sum_{k=0}^{r_0-1} \sum_{l=k+1}^{r_0} \dfrac{{(4e\beta)^{k+l}}}{(1-p)^k} \\
    & \leq \dfrac{4e\beta}{ 1 - 4e \beta} \dfrac{1 - p}{1 - p - 4e^2\beta^2}.
\end{align*}
Hence, $\sum_{r=0}^{r_0} \E[T^2 \mid R = r] \P[R=r]$ is uniformly bounded in $r_0$. Thus, $\E[\E(T^2 \mid R)] = \sum_{r=0}^\infty \E(T^2\mid R = r)\P(R = r) < \infty$.

\subsection{Proof of Theorem \ref{thm:noisy_sampler}}
Suppose $P(\theta, \cdot)$ and $\hat{P}_N(\theta, \cdot)$ denote the transition kernels resulting from $R_{MH}(\theta, \theta')$ and any noisy estimate $\hat{R}_N(\theta, \theta', u, u')$ where $u_N \sim F_\theta$ and $u_N' \sim F_{\theta'}$ are auxiliary variables drawn to create the estimate, and let us assume that $u$ and $u'$ are independent. Let $\norm{p -q }_{TV} = \int |p(x) - q(x)|dx$ denote the total variation distance between two densities $p,q$ with appropriate dominating measure. For the following result, the independence is not necessary, but it simplifies the calculation. 
A simple adaptation of Corollary 2.3 of \citet{alquier2016noisy} yields the following result:
$$\norm{P(\theta, \cdot) - \hat{P}_N(\theta, \cdot)}_{TV} \leq \sup_\theta \int \delta(\theta, \theta')q(\theta'\mid \theta) d\theta',$$
where, 
\begin{align*}
    \delta(\theta, \theta') = \E|\min\{1, R_{MH}(\theta, \theta')\} - \min\{1, \hat{R}_N(\theta, \theta', u_N, u_N')\}|,
\end{align*}
and the expectation is taken with respect to the product measure $F_\theta \times F_{\theta'}$. In other words, the total variation distance between the transition kernels depend on the quality of the approximation in expectation. In particular, if the data support is bounded, then one can get away by approximating $R_{MH}$ in the log-scale. This is crucial for numerical stability. Indeed, if $ a\leq X \leq b$. Then from the mean value theorem, it follows that, there exists $c_1$ and $c_2$ such that 
    $$c_1 \E|X - c| \leq \E|e^X - e^c | \leq c_2 \E|X - c|.$$

Next, note that  when $\Theta$ is a bounded subset of $\mathbb{R}^{p \times p}$, $z(\theta) \in [z_1, z_2]$. By a similar argument, $\pi(\theta)$ and $q(\cdot\mid \theta')$ are also bounded. If in addition, the support of the PEGM is bounded, which is true for the Ising model, then $V(\theta, \theta')$ is also bounded. Hence, by our previous discussion,
$$\E| V(\theta, \theta') - \log R_{MH}(\theta, \theta')| \asymp \E|e^{V(\theta, \theta')} - R_{MH}(\theta, \theta')|.$$
We now study the estimator $V(\theta, \theta')$. In the following calculations, all expectations are taken with respect to $F_\theta \times F_{\theta'}$. We have
\begin{align*}
    &\E|V(\theta, \theta') -\log R_{MH}(\theta, \theta')|\\
    & \leq n\E\left| \log \widetilde{T}(\theta) - \log \frac{z(\theta)}{z(\phi)}\right| +  n\E\left| \log \widetilde{T}(\theta') - \log \frac{z(\theta')}{z(\phi')}\right|\\
    & \asymp n \E\left| \widetilde{T}(\theta) - \frac{z(\theta)}{z(\phi)}\right| + n \E\left| \widetilde{T}(\theta') - \frac{z(\theta')}{z(\phi')}\right| \\
    & \leq n \sqrt{\text{var}(\widetilde{T}(\theta))} + n \sqrt{\text{var}(\widetilde{T}(\theta'))} = O(1/\sqrt{N}),
\end{align*}
since from Proposition \ref{prop:var_W}, $\text{var}(\widetilde{T}(\theta)) = N^{-1} [z(2\theta - \phi)/z(\phi) - z^2(\theta)/z^2(\phi)]$. We are now ready to prove the theorem.

Since $\pi(\theta)$ and $q(\cdot\mid \theta)$ are continuous over $\Theta$, they are bounded over $\Theta$. Also, $\sup_{\theta \in \Theta} \norm{\theta} \leq p B $. Let $\sup_\theta \pi(\theta)\leq c_\pi$ and $\sup_\theta q(\cdot\mid \theta) \leq c_q$. Hence, the first claim follows from Theorem 16.0.2 of \cite{meyn2012markov}, (see also Theorem 3.2 of \cite{alquier2016noisy}) with $C = 2$ and $\rho = 1 - 1/(c_\pi^3c_q^3 (pB)^4)$. The second claim also follows similarly from Theorem 3.2 of \cite{alquier2016noisy}.

\subsection{Proof of Corollary \ref{cor:cor1}}
By the triangle inequality,
\begin{align*}
    \norm{\delta_{\theta_0} \hat{P}_N^t - \pi(\cdot \mid \mathcal{D})}_{TV} &\leq \norm{\delta_{\theta_0} \hat{P}_N^t - \delta_{\theta_0}P^t}_{TV}  + \norm{\delta_{\theta_0} P^t - \pi(\cdot \mid \mathcal{D})}_{TV}.
\end{align*}
The result follows from Theorem \ref{thm:noisy_sampler} and letting $N\to \infty$.

\section{Constructing Gradient-based Proposals}\label{sec:approx_gradient}
First, note that $\nabla_\theta \log \pi(\theta \mid \mathcal{D}) = \sum_{l=1}^n \nabla_\theta f(\x_l;\theta) - n \nabla_\theta \log z(\theta) + \nabla_\theta \log \pi(\theta)$. The intractable term is $\nabla_\theta \log z(\theta)$. It is easily seen that $\nabla_\theta \log z(\theta) = \E[\nabla_\theta \log f(X;\theta)]$ where $X \sim p(\cdot;\theta)$. This motivates a Monte-Carlo estimate but the key issue is sampling $X \sim p(\cdot;\theta)$. To avoid this complication, we use the fact that $\nabla_\theta \log z(\theta) = \nabla_\theta z(\theta)/z(\theta)$. Next, under standard regularity conditions,
\begin{align*}
    \nabla_\theta z(\theta) = \nabla_\theta \int f(\x;\theta) dx &= \int \nabla_\theta f(\x;\theta)dx \\
    & = \int \dfrac{\nabla_\theta f(\x;\theta)}{p(\x;\phi)} p(\x;\phi)dx \\
    & = \E_{X \sim p(\cdot;\phi)}\left[ \dfrac{\nabla_\theta f(X;\theta)}{p(\x;\phi)}\right].
\end{align*}
Thus a Monte-Carlo estimate of $\nabla_\theta z(\theta)$ is:
$$ \widetilde{T}^{\nabla}(\theta)= \frac{1}{N} \sum_{i=1}^N \dfrac{\nabla_\theta f(\y_i;\theta)}{p(\y_i;\phi)}, \quad \y_i \overset{iid}{\sim} p(\cdot;\phi).$$
Finally, recalling the estimator of $\widetilde{T}(\theta)$ of $z(\theta)/z(\phi)$, a ratio estimator of $\nabla_\theta \log z(\theta)$ is $\frac{\widetilde{T}^\nabla(\theta)}{\widetilde{T}(\theta)}$.

\section{Additional Numerical Experiments}\label{sec:supp_num}

\subsection{Sensitivity to the importance sample size $N$}

Table~\ref{tab:var_run} reports the empirical variance of the unbiased estimator $\tilde{T}$ of $z(\theta)/z(\phi)$ and the corresponding runtime across 100 replications for the Ising model. As expected, increasing $N$ reduces the variance of $T$ across all dimensions, with reductions of roughly an order of magnitude for each tenfold increase in $N$. This improvement comes at a proportional increase in computational cost.

\begin{table}[!h]
    \centering
    \caption{Empirical variance of the unbiased estimator $\tilde{T}$ of $z(\theta)/z(\phi)$ and runtime (seconds) across 100 replications for Ising model.}
    \label{tab:var_run}
    \begin{tabular}{l cc cc cc}
        \toprule
        & \multicolumn{2}{c}{$N = 5000$} & \multicolumn{2}{c}{$N = 50000$} & \multicolumn{2}{c}{$N = 500000$} \\
        \cmidrule(lr){2-3} \cmidrule(lr){4-5} \cmidrule(lr){6-7}
        & $\mathrm{Var}(\tilde{T})$ & time (s) & $\mathrm{Var}(\tilde{T})$ & time (s) & $\mathrm{Var}(\tilde{T})$ & time (s) \\
        \midrule
        $p = 5$   & $3.15 \times 10^{-5}$  & 0.0008 & $3.20 \times 10^{-6}$  & 0.0079 & $3.09 \times 10^{-7}$  & 0.0795 \\
        $p = 50$  & $9.94 \times 10^{-7}$  & 0.0066 & $1.73 \times 10^{-7}$  & 0.0678 & $1.15 \times 10^{-8}$  & 0.6950 \\
        $p = 100$ & $1.25 \times 10^{-8}$  & 0.0149 & $1.09 \times 10^{-9}$  & 0.1523 & $1.33 \times 10^{-10}$ & 1.5188 \\
        \bottomrule
    \end{tabular}
\end{table}

\subsection{Effective sample size per unit time}
    \begin{table*}[h]
    \caption{ESS/minute for the PM, N, EX samplers with Langevin proposal.}
    \centering
    \begin{tabular}{l cc cc cc}
        \toprule
        & PM & N & EX \\
        \hline
        $p = 50$  &  7.53 & 16.35  & 14.52 \\
        $p = 70$  &  6.03 &  1.75  & 1.70  \\
        $p = 100$ &  3.05 &  0.88 & 0.97  \\
        \bottomrule
    \end{tabular}
    \label{tab:ess_min}
\end{table*}

To further illustrate the scalability of the proposed approach, Table \ref{tab:ess_min} reports the effective sample size per minute (ESS/minute) for the proposed pseudo-marginal sampler, noisy sampler, and the exchange algorithm with the Langevin proposal. At $p = 50$, the Noisy and the Exchange algorithm perform much better than the pseudo-marginal sampler. However, as the dimension increases beyond $p=50$, the performance of both the Noisy and the (approximate) exchange algorithm deteriorates. This is potentially due to the poor mixing of the inner Gibbs chain, whose computational cost is no longer offset by any gains in sampling efficiency at higher dimensions. In contrast, the pseudo-marginal sampler with the Langevin proposal performs better. 

To summarize, the pseudo-marginal sampler, while computationally expensive, provides better effective sample sizes. This is especially important in Bayesian inference since the ultimate goal of posterior sampling using MCMC is to approximate posterior expectations of various kinds. Having a larger effective sample size essentially contributes to lower variance estimators from the pseudo-marginal chain.
\section{Additional Data Analysis Results}\label{sec:supp_data}
\begin{figure}[!h]
    \centering
\includegraphics[height = 6cm, width = 6cm]{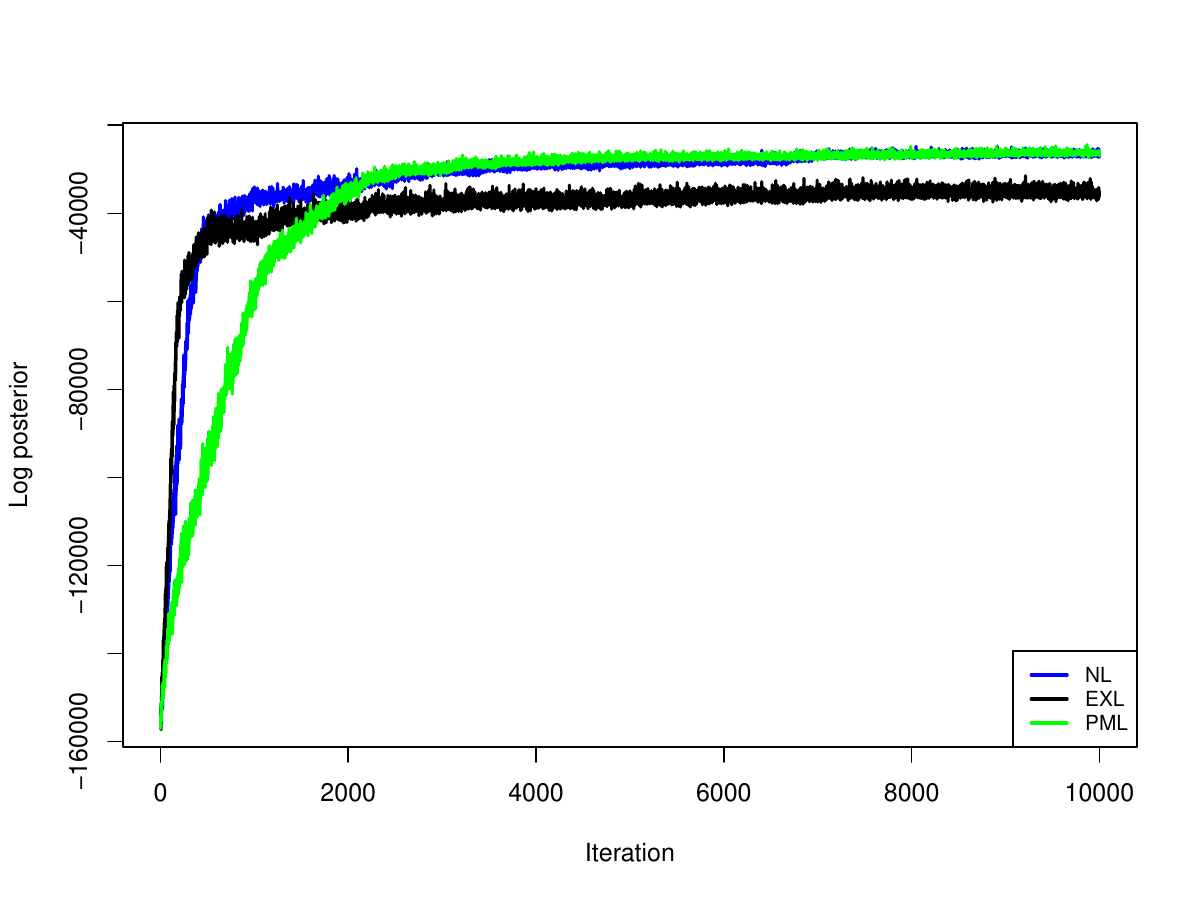}
    \caption{Log posterior trace plots for the PM, N, EX samplers with Langevin proposal for the movie data.}
    \label{fig:Movie_log_pos}
\end{figure}

\begin{figure}
    \centering
\includegraphics[width=\linewidth]{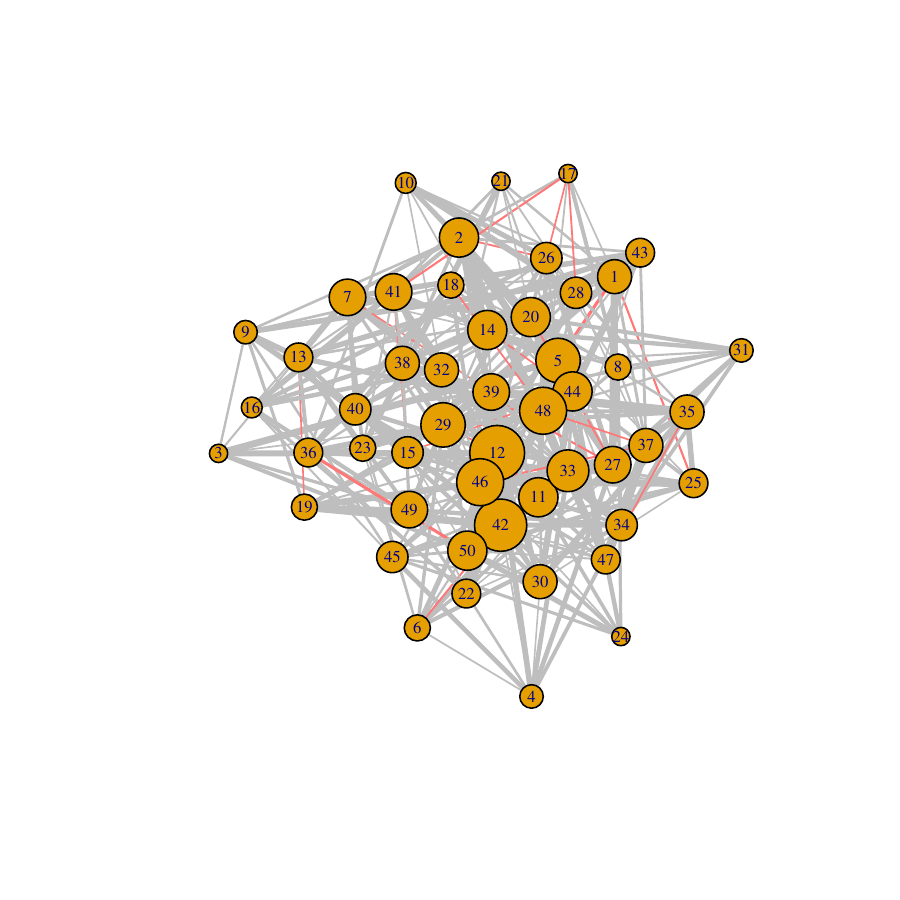} 
\caption{PM(L)-based Ising ($\theta^{50\times50}$) Model Movie Network.Thicker edges indicate higher absolute values of posterior mean estimates, while larger nodes represent higher degrees, and red versus gray distinguishes between shared and contrasting preferences.}
\label{fig:network_lmc}
\end{figure}

\begin{figure}
\centering
\includegraphics[width=\linewidth]{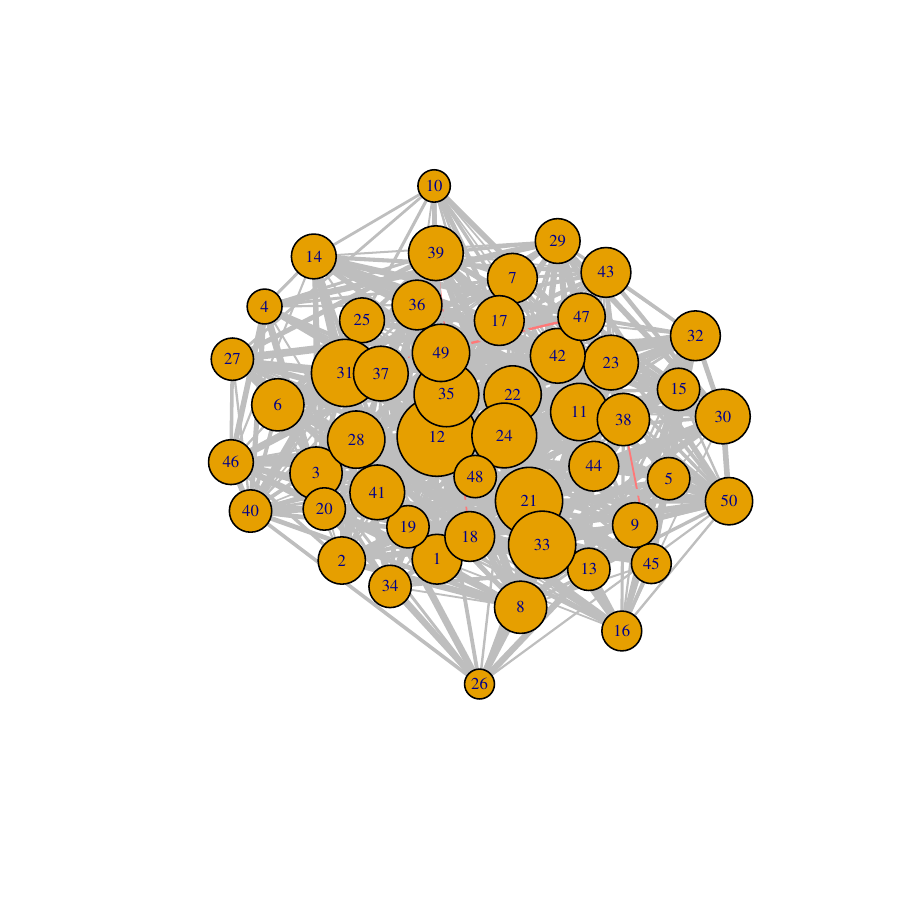} 
\caption{EX(L)-based Ising ($\theta^{50\times50}$) Model Movie Network.Thicker edges indicate higher absolute values of posterior mean estimates, while larger nodes represent higher degrees, and red versus gray distinguishes between shared and contrasting preferences.}
\label{fig:network_exlmc}
\end{figure}

\begin{figure}
\centering
\includegraphics[width=\linewidth]{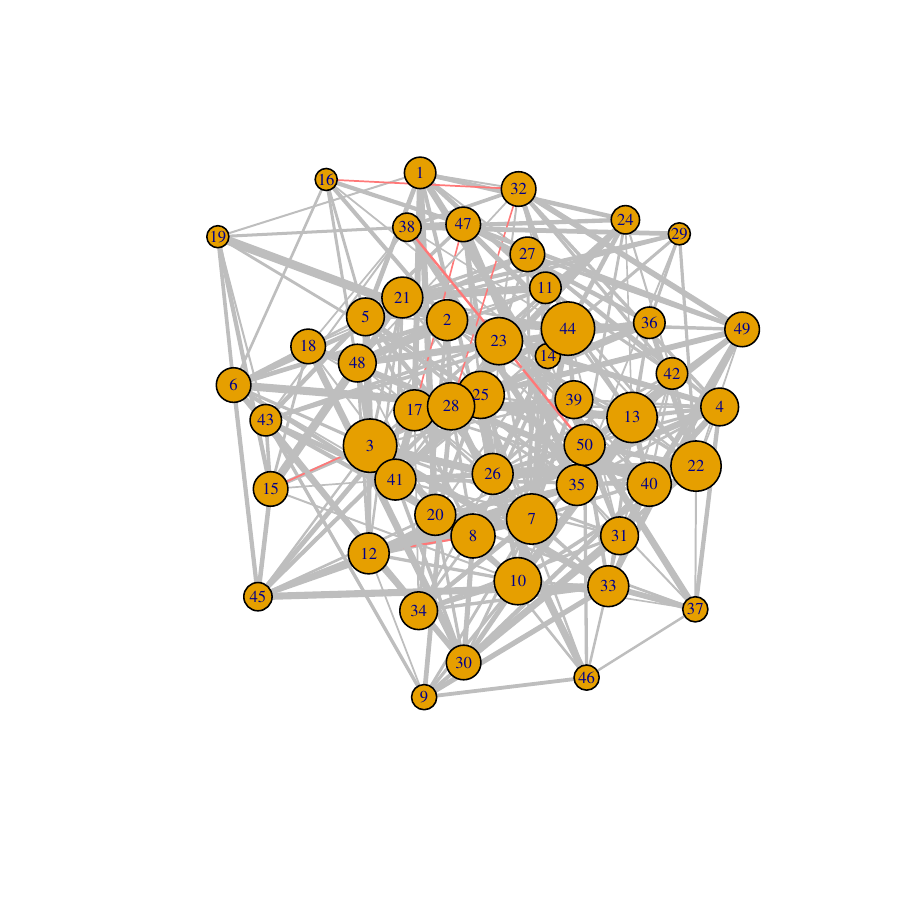} 
\caption{N(L)-based Ising ($\theta^{50\times50}$) Model Movie Network.Thicker edges indicate higher absolute values of posterior mean estimates, while larger nodes represent higher degrees, and red versus gray distinguishes between shared and contrasting preferences.}
\label{fig:network_nlmc}
\end{figure}

Figure \ref{fig:Movie_log_pos} shows that while the pseudo-marginal method converges slightly slower than both the noisy and exchange methods, it ultimately reaches log-likelihood values comparable to those of the noisy method. In contrast, the exchange method converges to noticeably lower log-posterior values.  
Tables \ref{tab:movie_exl},  \ref{tab:movie_pml}  and \ref{tab:movie_NL} show the ten strongest positive and negative connections (i.e., the largest and smallest $\hat{\theta}_{jk}$ values) from the posterior mean estimates of the parameter matrix $\theta$ for each method. Among the three, the pseudo-marginal approach produces the most interpretable results. Most of the identified connections align with intuitive expectations. For instance, the strong common preference between animated films \textit{The Lion King} and \textit{Toy Story}, and the opposite preference between the psychological thriller \textit{Memento} and the epic fantasy \textit{The Lord of the Rings}. The movie IDs are presented in Table~\ref{tab:movie_id}. Figures~\ref{fig:network_lmc}, \ref{fig:network_exlmc} and \ref{fig:network_nlmc} provide visualizations of the networks resulting under different methods.

\begin{table*}[h!]
    \centering
\caption{Top 10 positive and negative interactions - EX(L) Method}
\scalebox{0.65}{
\begin{tabular}{c|c||c|c}
\hline
\textbf{Positive Edge} & $\hat{\theta}_{jk}$ & \textbf{Negative Edge} & $\hat{\theta}_{jk}$\\

\hline
Gladiator (2000) - The Lord of the Rings (2002) & 0.15 & Blade Runner (1982) - Shrek (2001) & -0.68 \\
\hline
The Dark Knight (2008) - The Matrix (1999) & 0.12 & Shrek (2001) - Pirates of the Caribbean (2003) & -0.63 \\
\hline
Speed (1994) - Inception (2010) & 0.11 & Gladiator (2000) - Groundhog Day (1993) & -0.61 \\
\hline
Schindler's List (1993) - The Fugitive (1993) & 0.09 & Back to the Future (1985) - The Lord of the Rings (2001) & -0.59 \\
\hline
Monty Python and the Holy Grail (1975) - Men in Black (1997) & 0.09 & The Godfather  (1972) - The Matrix (1999) & -0.58 \\
\hline
Monty Python and the Holy Grail (1975) - Star Wars V (1980) & 0.09 & Star Wars VI  (1983) - Terminator 2 (1991) & -0.56 \\
\hline
Seven (1995) - Indiana Jones and the Last Crusade (1989) & 0.08 & The Shawshank Redemption (1994) - The Godfather (1972) & -0.56 \\
\hline
The Silence of the Lambs (1991) - Back to the Future (1985) & 0.08 & Gladiator (2000) - The Dark Knight (2008) & -0.55 \\
\hline
Pulp Fiction (1994) - The Godfather (1972) & 0.08 & Twelve Monkeys (1995) - Pirates of the Caribbean  (2003) & -0.54 \\
\hline
Forrest Gump (1994) - Star Wars VI (1983) & 0.08 & Terminator 2 (1991) - The Godfather  (1972) & -0.54 \\
\hline
\end{tabular}
}
\label{tab:movie_exl}
\end{table*}

\begin{table}[h!]
\centering
\caption{Top 10 positive and negative interactions - PM(L) Method}
\scalebox{0.75}{
\begin{tabular}{c|c||c|c}
\hline
\textbf{Positive Edge} & $\hat{\theta}_{jk}$ & \textbf{Negative Edge} & $\hat{\theta}_{jk}$\\
\hline
Twelve Monkeys  (1995) - Schindler's List (1993) & 0.20 & Memento (2000) - The Lord of the Rings (2002) & -0.62 \\
\hline
Fight Club (1999) - Groundhog Day (1993) & 0.20 & The Lord of the Rings (2001) - Inception (2010) & -0.53 \\
\hline
Twelve Monkeys  (1995) - Terminator 2 (1991) & 0.13 & The Silence of the Lambs (1991) - Seven (1995) & -0.52 \\
\hline
Independence Day  (1996) - Shrek (2001) & 0.13 & Braveheart (1995) - Saving Private Ryan (1998) & -0.50 \\
\hline
The Lion King (1994) - Toy Story (1995) & 0.13 & Braveheart (1995) - The Lord of the Rings (2002) & -0.49 \\
\hline
Blade Runner (1982) - The Terminator (1984) & 0.13 & Twelve Monkeys  (1995) - Independence Day  (1996) & -0.48 \\
\hline
Memento (2000) - The Sixth Sense (1999) & 0.12 & Batman (1989) - The Lord of the Rings (2003) & -0.48 \\
\hline
Pulp Fiction (1994) - Toy Story (1995) & 0.12 & True Lies (1994) - The Terminator (1984) & -0.48 \\
\hline
Star Wars VI  (1983) - Raiders of the Lost Ark (1981) & 0.12 & Terminator 2 (1991) - The Sixth Sense (1999) & -0.47 \\
\hline
Braveheart (1995) - Dances with Wolves (1990) & 0.12 & Braveheart (1995) - Independence Day (1996) & -0.46 \\
\hline
\end{tabular}
}
\label{tab:movie_pml}
\end{table}

\begin{table}[h!]
\centering
\caption{Top 10 positive and negative interactions - N(L) Method}
\scalebox{0.75}{
\begin{tabular}{c|c||c|c}
\hline
\textbf{Positive Edge} & $\hat{\theta}_{jk}$ & \textbf{Negative Edge} & $\hat{\theta}_{jk}$\\
\hline
The Dark Knight (2008) - Groundhog Day (1993) & 0.15 & Batman (1989) - Dances with Wolves (1990) & -0.68 \\
\hline
Fargo (1996) - Star Wars VI (1983) & 0.13 & Terminator 2 (1991) - Dances with Wolves (1990) & -0.67 \\
\hline
The Shawshank Redemption (1994) - Shrek (2001) & 0.11 & True Lies (1994) - Apollo 13 (1995) & -0.56 \\
\hline
Pulp Fiction (1994) - The Godfather (1972) & 0.11 & The Silence of the Lambs (1991) - Star Wars VI (1983) & -0.54 \\
\hline
The Lion King (1994) - The Lord of the Rings (2002) & 0.10 & Pulp Fiction (1994) - The Lord of the Rings (2003) & -0.54 \\
\hline
The Godfather (1972) - Shrek (2001) & 0.10 & Star Wars V (1980) - The Terminator (1984) & -0.53 \\
\hline
Twelve Monkeys (1995) - Star Wars IV (1977) & 0.10 & Blade Runner (1982) - Back to the Future (1985) & -0.52 \\
\hline
Braveheart (1995) - Independence Day  (1996) & 0.10 & Star Wars IV (1977) - Dances with Wolves (1990) & -0.50 \\
\hline
Star Wars VI (1983) - Batman (1989) & 0.09 & The Princess Bride (1987) - Aladdin (1992) & -0.49 \\
\hline
Saving Private Ryan (1998) - The Lion King (1994) & 0.09 & Braveheart (1995) - The Lord of the Rings (2003) & -0.49 \\
\hline
\end{tabular}
}
\label{tab:movie_NL}
\end{table}

\begin{table}[h!]
\centering
\caption{Top 50 ranked movies with movie ID, title, and genre classifications}
\scalebox{0.75}{
\begin{tabular}{c|c|c}
\hline
\textbf{Movie ID} &  \textbf{Title} & \textbf{Genres} \\
\hline
1 &  Twelve Monkeys (1995) & Mystery$|$Sci-Fi$|$Thriller \\
2 &  Braveheart (1995) & Action$|$Drama$|$War \\
3 &  Star Wars: Episode IV - A New Hope (1977) & Action$|$Adventure$|$Sci-Fi \\
4 &  Forrest Gump (1994) & Comedy$|$Drama$|$Romance$|$War \\
5 &  Schindler's List (1993) & Drama$|$War \\
6 &  Blade Runner (1982) & Action$|$Sci-Fi$|$Thriller \\
7 & The Silence of the Lambs (1991) & Crime$|$Horror$|$Thriller \\
8 & Fargo (1996) & Comedy$|$Crime$|$Drama$|$Thriller \\
9 & Monty Python and the Holy Grail (1975) & Adventure$|$Comedy$|$Fantasy \\
10 &  Star Wars: Episode V - The Empire Strikes Back (1980) & Action$|$Adventure$|$Sci-Fi \\
11 &  The Princess Bride (1987) & Action$|$Adventure$|$Comedy$|$Fantasy$|$Romance \\
12 &  Star Wars: Episode VI - Return of the Jedi (1983) & Action$|$Adventure$|$Sci-Fi \\
13 &  Back to the Future (1985) & Adventure$|$Comedy$|$Sci-Fi \\
14 &  Saving Private Ryan (1998) & Action$|$Drama$|$War \\
15 &  Pulp Fiction (1994) & Comedy$|$Crime$|$Drama$|$Thriller \\
16 & The Shawshank Redemption (1994) & Crime$|$Drama \\
17 & The Lion King (1994) & Adventure$|$Animation$|$Children$|$Drama$|$Musical$|$IMAX \\
18 &  Speed (1994) & Action$|$Romance$|$Thriller \\
19 &  True Lies (1994) & Action$|$Adventure$|$Comedy$|$Romance$|$Thriller \\
20 &  The Fugitive (1993) & Thriller \\
21 & Aladdin (1992) & Adventure$|$Animation$|$Children$|$Comedy$|$Musical \\
22 &  Batman (1989) & Action$|$Crime$|$Thriller \\
23 &  Apollo 13 (1995) & Adventure$|$Drama$|$IMAX \\
24 &  Jurassic Park (1993) & Action$|$Adventure$|$Sci-Fi$|$Thriller \\
25 &  Terminator 2: Judgment Day (1991) & Action$|$Sci-Fi \\
26 &  Dances with Wolves (1990) & Adventure$|$Drama$|$Western \\
27 &  Independence Day  (1996) & Action$|$Adventure$|$Sci-Fi$|$Thriller \\
28 & The Godfather (1972) & Crime$|$Drama \\
29 &  Raiders of the Lost Ark  (1981) & Action$|$Adventure \\
30 &  American Beauty (1999) & Drama$|$Romance \\
31 &  Gladiator (2000) & Action$|$Adventure$|$Drama \\
32 &  Shrek (2001) & Adventure$|$Animation$|$Children$|$Comedy$|$Fantasy|Romance \\
33 &  Pirates of the Caribbean: The Curse of the Black Pearl (2003) & Action$|$Adventure$|$Comedy$|$Fantasy \\
34 &  Seven  (1995) & Mystery$|$Thriller \\
35 &  The Lord of the Rings: The Fellowship of the Ring (2001) & Adventure$|$Fantasy \\
36 &  Fight Club (1999) & Action$|$Crime$|$Drama$|$Thriller \\
37 &  Memento (2000) & Mystery$|$Thriller \\
38 &  Dark Knight, The (2008) & Action$|$Crime$|$Drama$|$IMAX \\
39 &  Inception (2010) & Action$|$Crime$|$Drama$|$Mystery$|$Sci-Fi$|$Thriller$|$IMAX \\
40 &  The Usual Suspects (1995) & Crime$|$Mystery$|$Thriller \\
41 &  Toy Story (1995) & Adventure$|$Animation$|$Children$|$Comedy$|$Fantasy \\
42 &  The Terminator (1984) & Action$|$Sci-Fi$|$Thriller \\
43 &  Indiana Jones and the Last Crusade (1989) & Action$|$Adventure \\
44 &  Men in Black (1997) & Action$|$Comedy$|$Sci-Fi \\
45 &  The Matrix (1999) & Action$|$Sci-Fi$|$Thriller \\
46 &  The Sixth Sense (1999) & Drama$|$Horror$|$Mystery \\
47 &  The Lord of the Rings: The Two Towers (2002) & Adventure$|$Fantasy \\
48 &  The Lord of the Rings: The Return of the King (2003) & Action$|$Adventure$|$Drama$|$Fantasy \\
49 &  Good Will Hunting (1997) & Drama$|$Romance \\
50 &  Groundhog Day (1993) & Comedy$|$Fantasy$|$Romance \\
\hline
\end{tabular}
}
\label{tab:movie_id}
\end{table}

\clearpage

\clearpage
\end{document}